\documentclass[LectureNotes]{SciPost}
\pdfoutput=1

\usepackage{float}
\restylefloat{table}
\usepackage{microtype}
\usepackage{xcolor}
\usepackage[most]{tcolorbox}
\usepackage[utf8]{inputenc}
\usepackage{amsmath,amssymb,amsfonts,amsthm,mathtools,mathrsfs}
\usepackage{babel}
\usepackage{slashed}
\usepackage[sort&compress,numbers]{natbib}
\usepackage{tocloft}
\usepackage{svg}
\usepackage{tikz}
\usepackage{tikz-feynman}
\usepackage{feynmp-auto}

\newcommand{\D}{\mathcal{D}}
\definecolor{yellow1}{HTML}{ffffcc}
\definecolor{myviolet}{HTML}{007B8B}
\definecolor{mylavender}{HTML}{DEFEFF}
\definecolor{soft}{HTML}{fff6ed}
\definecolor{sky}{HTML}{CEFFFF}
\definecolor{indigo}{HTML}{000066}
\definecolor{myblue}{HTML}{056FB1}
\definecolor{maroon}{HTML}{db0000}
\definecolor{forest}{HTML}{06961c}
\definecolor{lime}{HTML}{fcffad}
\definecolor{mygreen}{HTML}{024f16}
\definecolor{newgreen}{HTML}{aaf50a}
\definecolor{ocre}{HTML}{eb6315}

\usepackage{hyperref}
\usepackage{prettyref}
\newcommand{\pref}{\prettyref}
\newrefformat{eq}{equation (\ref{#1})}
\newrefformat{sec}{section~\ref{#1}}
\usepackage{bbold}

\newtcolorbox[auto counter,number within=section]{example}[2][]{ title style={right color=myblue!20, left color=myblue},
	colback=mylavender,colframe=newgreen!5,interior style={left color=newgreen!50,right color=white},fonttitle=\bfseries,breakable,enhanced jigsaw,
	title=Remark~\thetcbcounter: #2,#1}

\newtcolorbox{side}[1]{colback=lime,
	colframe=lime!20,interior style={left color=lime,right color=white},breakable,enhanced jigsaw}
\newtcolorbox{pozor}[1]{enhanced,colback=white,
	colframe=soft!20, interior style={left color=red!15,right color=white},fonttitle=\bfseries,breakable,enhanced jigsaw,
	title=#1}
\newtcolorbox[]{definition}[1][]{enhanced,borderline west={3pt}{-3pt}{cyan},colback=white,colframe =white, interior style={left color = ocre!5, right color =white}, breakable,enhanced jigsaw}

\renewcommand{\d}{\mathrm{d}}
\hypersetup{
	colorlinks=true,
	linkcolor=ocre,
	filecolor=cyan,      
	citecolor=ocre,
    urlcolor=myblue
}
\usepackage{tensor}
\usepackage{comment}
\usepackage{parskip} 
\usepackage{mathrsfs}
\usepackage{cleveref}

\begin{document}
\begin{samepage}
		\begin{flushleft}\huge{\textbf{Generalised Symmetries and Manifest Duality I: Flat Spacetime}}\end{flushleft}
		\vspace{20pt}
		{\color{myviolet}\hrule height 1mm}
		\vspace*{10pt}
		\begin{flushleft}
		\large 	\textbf{Subhroneel Chakrabarti}$\,{}^a$,  \textbf{Arkajyoti Manna}$\,{}^{b,c}$, \textbf{Madhusudhan Raman}$\,{}^d$
		\end{flushleft}

		\begin{flushleft}
			\emph{\large ${}^a$ Department of Theoretical Physics and Astrophysics, Faculty of Science, Masaryk University, 611 37 Brno, Czech Republic.}
			\\ \vspace{1mm}
            \emph{\large ${}^b$ The Institute of Mathematical Sciences \\
 	IV Cross Road, C.I.T. Campus, Taramani, Chennai 600 113, India}
			\\ \vspace{1mm}
            \emph{\large ${}^c$ Homi Bhabha National Institute \\
	Training School Complex, Anushakti Nagar, Mumbai 400 094, India}
			\\ \vspace{1mm}
            \large ${}^d$ \emph{Department of Physics and Astrophysics\\
            University of Delhi, Delhi 110 007, India}
			\\ \vspace{3mm}
             \href{mailto:subhroneelc@sci.muni.cz}{subhroneelc@sci.muni.cz},
             \href{mailto:akm.manna@gmail.com}{akm.manna@gmail.com},
             \href{mailto:madhusudhan.raman@gmail.com}{madhusudhan.raman@gmail.com} \\
		\end{flushleft}

  \begin{flushright}
		\end{flushright}
		\section*{Abstract}
		{
			We present a novel, manifestly Lorentz‑invariant, local, polynomial, and straightforwardly quantisable action for duality‑symmetric gauge theories formulated using gauge potentials and fluxes. This new action possesses a novel higher-form gauge symmetry that we dub $h$-gauge symmetry, which on gauge fixing can lead to Sen's formalism or to a solely potential-based description. Unlike Sen's flux‑based formalism, the potential-based action admits a simple minimal coupling to matter, thereby allowing us perform a number of consistency checks, including incorporating dynamical matter and establishing the Witten effect. The new action admits remarkably simple supersymmetrisation. We demonstrate the formalism explicitly for quantum electro‑magnetodynamics in $D=4$. We also present a proof of charge quantisation that applies equally to our and Sen’s formalisms. 
            }
\end{samepage}
	\newpage
	\vspace{10pt}
	\noindent\rule{\textwidth}{1pt}
	\pagecolor{white}
    \setcounter{tocdepth}{1}
	\tableofcontents\thispagestyle{fancy}
	\noindent\rule{\textwidth}{1pt}
	\vspace{10pt}

\section{Introduction} \label{sec:intro}

Gauge theory actions with gauge potentials as dynamical variables almost always assume that the physical, gauge-invariant field strength is given by the curvature of the gauge potential. This obvious and automatic choice has two important consequences. First, the Bianchi identity is always true, even off-shell. Second, even in absence of sources, the free action has a preference of one description over the other; for example, in standard electrodynamics we have an ``electric'' description. These consequences are not a cause of concern, and indeed are helpful, when one works with matter sources which break electric-magnetic duality symmetry. This is not always the case. Theories containing both electrically and magnetically charged objects are ubiquitous in the low-energy sector of string theory. The problem is more acute for self-dual field strengths (in appropriate dimensions) where the Bianchi identity gets mapped to an equation of motion which is supposed to be true only \textit{on-shell}.

These issues usually manifest in the failure of Maxwell-like actions to adequately account for the dynamics of duality-symmetric theories. There have been numerous attempts to resolve these long-standing issues, most of them resulting in actions that fail to preserve at least one of the following desirable features: manifest Lorentz symmetry, locality, or a polynomial action that can be straightforwardly quantised using time-tested methods. The only formalism that preserves all of the above features known till date is due to Sen \citep{Sen:2015nph,Sen:2019qit}.\footnote{See \citep{Evnin:2022kqn} and references therein for a review of the previously considered approaches.} Sen's formalism has been studied extensively in various dimensions and has passed a number of non-trivial checks, especially involving quantum computations \citep{Lambert:2019diy,Andriolo:2020ykk,Chakrabarti:2020dhv,Chakrabarti:2022lnn,Lambert:2023qgs,Chakrabarti:2023czz,Vanichchapongjaroen:2025psm,Janaun:2024wya,Hull:2025rxy,Hull:2025bqo}.

Sen's formalism has three unusual features that, at first sight, seems unwieldy. However, as has been recently shown \citep{Mamade:2025jbs} (also see \citep{Sen:2015nph}), this action is indeed what is obtained as the low-energy effective field theory starting from superstring field theory. These strange but stringy features are as follows:
\begin{itemize}
    \item The action contains an additional ``shadow'' sector, whose presence is required to write down a manifestly Lorentz invariant action. This sector nevertheless completely decouples from the Hamiltonian of the physical sector and therefore has no influence on the dynamics of the duality invariant physical sector.

    \item The action has a non-standard coupling to dynamical gravity. This is essential to ensure that the shadow sector does not couple to physical fields, including gravity.

    \item The action treats the flux as the dynamical variable and it does \emph{not} consider the flux as a field strength of a gauge potential. In other words the formalism does not invoke gauge potentials in anyway.
\end{itemize}

All of these features are expected from string (field) theory considerations. However, these non-standard features certainly pose challenges that must be overcome. The decoupling of the shadow sector is rigorously proved using either a Hamiltonian analysis or using non-local, but well-defined, field-redefinitions in the momentum-space action. The non-standard coupling to gravity manifests in exotic transformation properties under diffeomorphism; nevertheless, a precise prescription to construct field representations that transform as tensors under diffeomorphism exists. This rigorously demonstrates that the action is invariant under diffeomorphism. That the flux is the fundamental variable necessitates abandoning traditional sources (that might have ordinarily coupled to gauge fields) and instead requires that we consider sources that have one higher rank. Even with such a modification, it is quite remarkable that one can compute current-current scattering and obtain a number of non-trivial quantum checks for this formalism \citep{Aggarwal:2025fiq}.

Nevertheless, working with fluxes directly also implies that writing down explicit matter coupling involving the usual currents is, at best, obfuscated. It is not so much that this cannot be done; rather, our intuition derived from ``minimal coupling'' prescriptions involving gauge fields are in this case rendered useless. Relatedly, there are classic results, such as the Witten effect, that are not easily captured by discarding the gauge potentials altogether and working with only fluxes. 

In this article, we propose a new action that is inspired by Sen's action, but can be seen as a generalisation of it, in the following sense. Our action describes the self-dual fluxes redundantly, with one flux-like variable which can be written in terms of a gauge potential in such a way that only its self-dual part of the curvature enters the action. The redundancy is taken care of by a higher-form gauge symmetry that is novel. This gauge symmetry can be gauge fixed to yield either Sen's action (by gauge fixing the potential part to zero) or an equivalent action involving only the gauge potential (by setting the flux-like variable to zero). The advantage of the latter is in this iteration we need only work with gauge potentials, whose coupling to sources is more familiar. The resulting action has a higher-form global symmetry, that we dub harmonic higher-form symmetry, which is a residual of the higher-form gauge symmetry of the parent action that was gauge fixed. 

Our proposed action is identical to Sen's, which can be obtained as a gauge-fixed version of the novel action. However, the advantage of the new action is that it allows a different gauge fixing, where we can work with a gauge potential, instead of treating a flux-like field as a dynamical object. This innocuous change however has significant consequences:
\begin{itemize}
    \item The potential-based formalism inherits all of the advantages of Sen's --- manifest Lorentz invariance, locality, polynomiality, and a straightforwardly quantisable action.
    \item The decoupling of the shadow sector can now be achieved completely at the level of coordinate space action.
    \item The reinstatement of gauge fields means we can easily recover all the advantages of having a gauge field, such as minimal coupling to matter, the Witten effect, etc.
    \item The relation of our action and Sen's is transparent: Sen's action can be interpreted as a partial gauge fixing of our formalism. Consequently, \textit{all} perturbative quantum checks done using Sen's formalism are automatically subsumed in ours.
\end{itemize}

\begin{figure}[hbp!]
\centering
\includegraphics[width=0.75\textwidth]{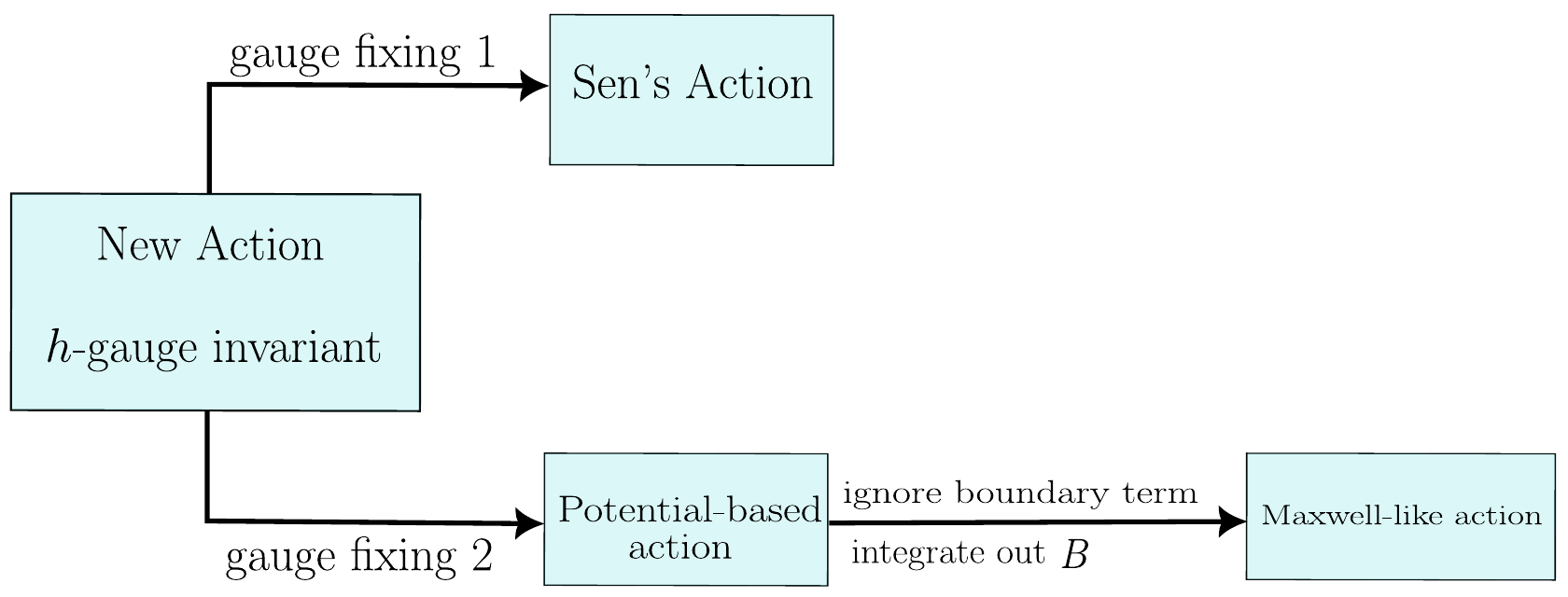}
\caption{The relationships between the new formalism, Sen's original formalism for self-dual form fields, and the potential-based formalism.}
\label{fig:map}
\end{figure}

Given the achievement of decoupling in co-ordinate space action, it might seem superfluous to keep on using the shadow sector to express the action. There are two reasons not to do that. First, the decoupling occurs \textit{only} for a specific choice of gauge fixing. For other equivalent choices, one needs to resort to Hamiltonian analysis, just like in Sen's formalism. Indeed, one ought to demonstrate, as we will, that the decoupling occurs for the parent theory.  Second, when coupling to gravity, one cannot simply carry out minimal coupling from the physical sector alone. The gravitational coupling of duality invariant theories still follows a non-standard route \textit{à la} Sen for which the shadow sector is essential. This is to be expected since it has been long known that S-duality constraints are incompatible with duality invariant actions coupled minimally to gravity under standard Kaluza-Klein reductions \citep{Witten:2009at,Lambert:2019diy}. In this paper we focus solely on flat spacetime and in a follow-up paper \citep{future} we detail the coupling to dynamical gravity (but see some comments in \pref{sec:disc}). 

The structure of the remainder of this paper is as follows. In \pref{sec:new_SD} we introduce our formalism for the self-dual case. We discuss the higher-form gauge symmetry and elaborate on how our action can be gauge fixed to Sen's formalism. In \pref{sec:Hamilton} we perform a rigorous Dirac constraint analysis of our theory and show the decoupling of the shadow sector at the level of Hamiltonian. Additionally, the constraint analysis makes it clear that our theory contains only self-dual degrees of freedom for the physical sector and just like Sen's, the anti-self-dual degrees of freedom do not enter at any point. Then we use our formalism for the case of electromagnetic duality in four dimensions in \pref{sec:new_4D}. We show how our formalism could be thought of as a two potential formalism, but unlike \citep{Zwanziger:1970hk} our action is perfectly Lorentz invariant and does not require any Dirac strings. In this section we also present a proof of Dirac's charge quantisation that holds equally for both ours and Sen's formalism. We discuss how the Witten effect is realised in our formalism as well. In \pref{sec:new_genD} we extend our action to discuss electric-magnetic duality in general dimensions for Abelian $p$-form theories with sources. Following that, in \pref{sec:susy} we discuss how to supersymmetrise our action. In contrast to Sen's formalism, supersymmetrisation in the potential-based action is much more straightforward and will hold at the level of action itself. In \pref{sec:example} we give an explicit coupling of our gauge theory in four dimensions to electrically and magnetically charged matter sectors. We show how the Feynman rules are derived and how they connect to those derived in \citep{Aggarwal:2025fiq} following Sen's formalism. We show how all computations in Sen's formalism are subsumed and superseded by our approach. We also present an explicit supersymmetrisation of the same action in four dimensions. We conclude in \pref{sec:disc} with some discussions of possible extensions and few comments on coupling to dynamical gravity.

\section{The New Action: Self-Dual Case}\label{sec:new_SD}
 Before we describe the new action, it is worthwhile to quickly recall Sen's formalism. We will work in $(4n+2)$-dimensional Minkowski spacetime where the principal physical field of interest is a self-dual $(2n+1)$-form.

 \subsection{Review of Sen's Formalism}
 Let us quickly review Sen's formalism (for more details we refer the readers to the original papers \citep{Sen:2015nph,Sen:2019qit}). The action reads

 \begin{equation} \label{eq:Sen}
     S_{_\mathrm{Sen}}[B,\mathcal{F}] = \frac{1}{2} \int \d B \wedge \star \, \d B - \int \d B \wedge \mathcal{F} + \frac{1}{2}\int \mathcal{F} \wedge \Omega \;, 
 \end{equation}
where $2n$-form fields $B$ are extra ``shadow'' degrees of freedom that completely decouple even at the quantum level, the self-dual flux $\mathcal{F}$ is treated as the dynamical variable that is \emph{not} thought of as the curvature of some gauge potential. The anti-self-dual $(2n+1)$-form $\Omega$ is the classical source for the self-dual flux.

The equations of motion are
\begin{align}
    &\delta_B S = 0 \implies \d \star \d B = \d \mathcal{F} \ , \\
    &\delta_\mathcal{F} S = 0 \implies \d B - \star \d B + \Omega = 0 \ .
\end{align}
In terms of the combination $\mathfrak{G} = \d B + \star \d B - \mathcal{F}$, the above equations of motion are equivalent to
\begin{align}
    &\d \star \mathfrak{G} = \d \mathfrak{G} = 0 \ , \\
    &\d \mathcal{F} = \d \star \mathcal{F} =\d \Omega \;.
\end{align}
In the above equations, the former is the decoupled extra degree of freedom which we can disregard. The latter, on identifying the source $\d\Omega = \star J$ where $J$ is the usual current $2n$-form, is precisely the desired equation of motion of a self-dual field-strength in presence of a classical source. \\

It was proved in \citep{Sen:2019qit} using the Hamiltonian analysis that the decoupling of the shadow sector holds exactly even in full interacting quantum theory, as long as the interaction term does not depend on the shadow sector. This decoupling, at the perturbative level, can also be seen at the level of momentum-space action after suitable field redefinitions \citep{Sen:2015nph,Chakrabarti:2022lnn,Aggarwal:2025fiq}.

\subsection{The New Action{}} \label{sec:Hamilton}

The dynamical fields are a pair of $2n$-forms $A$ and $B$, together with a self-dual $(2n+1)$-form $\Sigma$,
\begin{equation}
\Sigma := \star \Sigma .    
\end{equation}
We will also allow for possible interactions with other matter fields which we collectively denote by $\phi$. In what follows we will not write down the kinetic term for $\phi$ explicitly nor worry about its equation of motion. These will not affect any of the ensuing discussions.
The action is
\begin{equation}
S[B,A,\Sigma,\phi]
=
\frac12 \int_M \d B \wedge \star \d B
-
\int_M \d B \wedge F
+
S_i[F,\phi] +S_{\mathrm{kin}}[\phi] \,,
\label{gen-action}
\end{equation}
where
\begin{equation}
F := \d A + \star \d A - \Sigma ,
\label{def-F-general}
\end{equation}
and the interaction term $S_i$ depends on $A$ and $\Sigma$ only through the combination $F$, and will also depend on the additional matter fields $\phi$.  By construction, the interaction term does not depend on the extra field $B$. Note that the $F$ is also self-dual by definition and that this is true \textit{off-shell}. Exactly as in Sen's formalism, the field $B$ is an extra field which we will show decouple completely from the physical sector.  

Despite superficial similarity, our action is in fact a generalisation of Sen's action. Indeed, we have additional (higher-form) gauge symmetries, and naively we have two independent fields, viz., $A$ and $\Sigma$ that build the relevant self-dual flux $F$. However, as we will show later, the true physical degrees of freedom are indeed that of only a self-dual flux and there are no additional degrees of freedom in the physical sector.

\subsection{Equations of Motion}

Let us consider the interaction term to be just a source term corresponding to an anti-self-dual $(2n+1)$-form $\Omega$ that we treat as a background field. We have
\begin{equation}
    S_i = -\int F \wedge \Omega \;.
\end{equation}
The equations of motions now read
\begin{align}
    &\delta_B S = 0 \implies \d \star \d B = \d F \\
    &\delta_A S = 0 \implies \d\star \d B - \d \Omega = 0 \\
    & \delta_\Sigma S = 0 \implies \d B - \star \d B + \Omega = 0 \;.    
\end{align}
Note that the $\Sigma$ and $A$ equations of motion merely differ by an exterior derivative, so the $\Sigma$ equations of motion do not carry any new dynamical information. This is already indicative, at least on-shell, that there is a redundancy in description of $F$ in terms of the potential-like variable $A$ and the flux-like variable $\Sigma$.

Just like in Sen's formalism, in terms of the combination $\mathfrak{G} = \d B + \star \d B - F$, we obtain the desired decoupled equations of motion:
\begin{align}
    &\d \star \mathfrak{G} = \d \mathfrak{G} = 0 \ , \\
    &\d F = \d \star F = \d \Omega \;.
\end{align}
Once again, on identifying the source $\d\Omega = \star J$ where $J$ is the usual current $2n$-form we see these are the desired equations of motion for the physical sector. The new action, just like Sen's formalism, derives them by keeping Lorentz symmetry and self-duality manifest.

We have only established the decoupling of the shadow sector for the new action at the level of equations of motion here. To demonstrate its decoupling at the level of full quantum theory, we need to do a Hamiltonian analysis. Before proceeding to that, let us quickly see the various (gauge) symmetries that the new action possesses.

\subsection{The Higher-Form and Gauge symmetries}
The new action, in addition to manifest self-duality and Lorentz symmetry, posses a novel higher-form gauge symmetry that we dub $h$-gauge symmetry. The transformation of various fields under this is defined as

\begin{equation}
\delta_h B = 0 \quad , \quad \delta_h \phi = 0 \quad , \quad
\delta_h A = d^\dagger h \quad , \quad
\delta_h \Sigma = \Delta h \ ,
\end{equation}
with the gauge parameter $h$ a self-dual but otherwise arbitrary $(2n+1)$-form. We have denoted the adjoint of exterior derivative as $d^\dagger = \star \d \star$ and the Hodge Laplacian is denoted by $\Delta$. It is easy to see the combination $F$, indeed the action, is invariant under this $h$-gauge transformation.  The Hamiltonian analysis below will show that this redundancy appears as a first-class constraint, and that the physical sector, upon gauge fixing, is carried entirely by the self-dual field $F$. This gauge symmetry, upon being gauge-fixed offers us an option to express $F$ either only in terms of $\Sigma$ or the self-dual combination $(\d A + \star \d A)$. The former leads us back the Sen's action, whereas the latter will lead us to the Maxwell-like action with additional constraint, that is completely equivalent to Sen's formalism, but offers several distinct advantages that we will elaborate on due course.

The new action possess a few additional (gauge) symmetries. One of them is a higher-form symmetry and already present in Sen's formalism (although it is not usually identified as a higher-form symmetry) and requires the sources to be turned on; the other are familiar $U(1)$ gauge symmetries.

Firstly, note that there are two different $\mathrm{U}(1)$ gauge symmetries in the action. One for the extra fields $B$ and one for the gauge potentials $A$. The latter is not present in Sen's formalism and is unique to the new formalism.
\begin{equation}
    \delta B = \d \lambda_1 \quad, \quad \delta A = \d \lambda_2 \;,
\end{equation}
for two $(2n-1)$-form gauge parameters $\lambda_{1,2}$. 

The higher-form gauge symmetry that is familiar from Sen's formalism is only present in presence of sources. For the action
\begin{equation}
    S = \frac{1}{2} \int \d B \wedge \star \, \d B - \int \d B \wedge F  - \int F \wedge \Omega \;, 
\end{equation}
the higher-form gauge symmetry is parametrised by a $(2n)$-form parameter $\Lambda$ and it is defined as follows
\begin{align}
    \delta_\Lambda B &= \Lambda \;, \\
    \delta_\Lambda A &= \frac1{2}\Lambda \;, \, \delta_{\Lambda} \Sigma = -\frac1{2}(\d \Lambda + \star\, \d \Lambda) \;   \implies \delta F = \d \Lambda + \star\, \d \Lambda \\
    \delta_\Lambda \Omega &=  \d \Lambda   \;,
\end{align}
under which it is easily checked that the action is invariant (up to boundary terms) provided the parameter $\Lambda$ satisfies
    \begin{equation}
        \int \Lambda \wedge \d \Omega = 0 \;.
    \end{equation}
This gauge symmetry is directly analogous to the one already noted in Sen's formalism \citep{Sen:2015nph,Andriolo:2020ykk}. In the quantum theory, the action does not need to be exactly invariant, and one can relax the condition on $\Lambda$ to 
\begin{equation}
    \int \Lambda \wedge \star J = 2 \pi \mathbb{Z} \;.
\end{equation}
This higher-form symmetry was introduced in the original papers to capture the case of Type IIB SUGRA where the self-dual RR $5$-form flux couples to other form fields via Chern-Simons term that has a typical transformation property under the gauge transformation of the Kalb-Ramond and the RR $2$-form potential. However, we will see that this higher-form symmetry can also be interpreted as the generalised symmetry based interpretation of ``Dirac veto'' \citep{Hull:2024uwz} when one dimensionally reduces this theory over a torus to $4n$-dimensions to obtain an electric-magnetic duality invariant theory.

In summary, our new action subsumes all the (gauge) symmetries of Sen's formalism, while introducing additional ones that will be essential in ensuring that no new degrees of freedom are introduced. This result, along with the decoupling of the shadow sector, is most rigorously established in the Hamiltonian framework, to which we now turn.

\section{Hamiltonian Analysis}

The goal of this section is to perform a Hamiltonian analysis following Dirac-Bergmann procedure to explicitly establish two facts.
\begin{itemize}
    \item The shadow sector Hamiltonian decouples completely from the physical sector Hamiltonian, just like in Sen's formalism.
    \item The physical sector only contains degrees of freedom corresponding to a single self-dual $(2n+1)$-form field.
\end{itemize}

Readers who are only interested in the result may skip most of this section and only read \pref{sec:summ_H} for a summary. The following sections, while make use of the results, are independent of the details of the Hamiltonian analysis.

It will be useful to denote by $X$ the $(4n+1)$-dimensional flat spatial manifold with Euclidean flat metric. The spacetime Hodge star will always be denoted by $\star$, and we will also need the the Hodge star on $X$ during the Hamiltonian analysis, which we will denote by $\star_s$.

To keep the discussion parallel to Sen's treatment \citep{Sen:2019qit}, we fix the $U(1)$ gauge symmetries of $A$ and $B$ by choosing temporal gauge for both,
\begin{equation}
A_{0 i_1 \cdots i_{2n-1}} = 0 \quad ,
\quad
B_{0 i_1 \cdots i_{2n-1}} = 0 \ .
\end{equation}
After this gauge choice, we continue to denote the spatial $2n$-forms by the same symbols $A$ and $B$.

For a spatial $p$-form $\omega$ on $X$, the spacetime Hodge star acts as
\begin{equation}
\star (\d t \wedge \omega) = - \star_s \omega,
\qquad
\star \omega = (-1)^p \, \d t \wedge \star_s \omega .
\label{star-decomp}
\end{equation}
Since $X$ has dimension $4n+1$ and it is Euclidean, one has
\begin{equation}
\star_s^2 = 1
\end{equation}
on spatial $2n$-forms and on spatial $(2n+1)$-forms.

A self-dual $(2n+1)$-form can therefore be written uniquely as
\begin{equation}
\Sigma = \d t \wedge \sigma - \star_s \sigma ,
\label{Sigma-decomp}
\end{equation}
where $\sigma$ is a spatial $2n$-form.  Next, define the spatial $(2n)$-form
\begin{equation}
H_B := \star_s \d B ,
\qquad
H_A := \star_s \d A .
\label{def-H}
\end{equation}
Using \eqref{star-decomp}, one finds
\begin{equation}
\d A + \star \d A
=
\d t \wedge (\dot A - H_A) - \star_s (\dot A - H_A) .
\end{equation}
It follows that
\begin{equation}
F = \d t \wedge q - \star_s q \quad \mathrm{with}
\quad
q := \dot A - H_A - \sigma .
\label{eq:Fq-decomp}
\end{equation}
Thus the whole self-dual field $F$ is encoded in the single spatial $2n$-form $q$.  Conversely,
\begin{equation}
\star F
=
\star \bigl(\d t \wedge q - \star_s q \bigr)
=
- \star_s q + \d t \wedge q
=
F ,
\end{equation}
so self-duality is automatic.

\subsection{Lagrangians and Primary Constraints}

Using
\begin{equation}
\d B = \d t \wedge \dot B + \d B ,
\end{equation}
where the $\d B$ on RHS is understood to be the exterior derivative on $X$,
the first two terms in the action become
\begin{equation}
\frac12 \int_M \d B \wedge \star \d B
=
\int \d t \,
\Bigl[
-\frac12 \int_X \dot B \wedge \star_s \dot B
+\frac12 \int_X \d B \wedge \star_s \d B
\Bigr] ,
\end{equation}
and
\begin{equation}
-\int_M \d B \wedge F
=
\int \d t \,
\Bigl[
\int_X \dot B \wedge \star_s q
+
\int_X \d B \wedge q
\Bigr] .
\end{equation}
It is convenient to introduce the spatial inner product
\begin{equation}
\langle u , v \rangle := \int_X u \wedge \star_s v
\label{spatial-inner}
\end{equation}
for spatial $2n$-forms.  Since
\begin{equation}
\int_X \d B \wedge q = \langle H_B , q \rangle ,
\end{equation}
the Lagrangian takes the form
\begin{equation}
L
= \int \mathrm{d}t \left[ 
-\frac12 \langle \dot B , \dot B \rangle
+\frac12 \langle H_B , H_B \rangle
+\langle \dot B + H_B , q \rangle
+L_i[q,\phi] \right] \ .
\label{lagrangian-canonical}
\end{equation}
The interaction term $L_i$ is written as a functional of $q$ and the matter fields, since $F$ is completely equivalent to $q$ through \eqref{eq:Fq-decomp}.

The momenta conjugate to $B$, $A$, and $\sigma$ are
\begin{equation}
\Pi_B = \frac{\delta L}{\delta \dot B}
      = - \dot B + q ,
\label{PiB-general}
\end{equation}
\begin{equation}
\Pi_A = \frac{\delta L}{\delta \dot A}
      = \dot B + H_B + \frac{\delta L_i}{\delta q} ,
\label{PiA-general}
\end{equation} 
\begin{equation}
\Pi_\sigma = \frac{\delta L}{\delta \dot \sigma} = 0 .
\label{Pisigma-general}
\end{equation}
Thus
\begin{equation}
\Pi_\sigma \approx 0
\label{primary-general}
\end{equation}
is a primary constraint.

Combining \eqref{PiB-general} and \eqref{PiA-general}, we obtain
\begin{equation}
Y := \Pi_A + \Pi_B - H_B
=
q + \frac{\delta L_i}{\delta q} .
\label{Y-def}
\end{equation}
It is useful to regard it as the derivative of the functional
\begin{equation}
U[q,\phi] := \int \mathrm{d}t \left[  \frac12 \langle q,q\rangle + L_i[q,\phi] \right] \quad \mathrm{with}
\quad
Y = \frac{\delta U}{\delta q} .
\label{U-def}
\end{equation}
We now assume the standard non-degeneracy condition that this relation can be inverted, so that $q$ can be expressed in terms of $Y$ and the matter variables.

\subsection{Hamiltonian and Secondary Constraints}

Let $\Pi_\phi$ denote the canonical momenta conjugate to the matter fields.  We define the partial Legendre transform
\begin{equation}
G[Y,\phi,\Pi_\phi]
:=
\langle Y,q\rangle
+
\int_X \Pi_\phi \, \dot \phi
-
\Bigl(
\frac12 \langle q,q\rangle + L_i[q,\phi]
\Bigr) ,
\label{G-def}
\end{equation}
where on the right-hand side one expresses $q$ and the matter velocities in terms of $Y$, $\phi$, and $\Pi_\phi$.

A straightforward calculation then gives the canonical Hamiltonian
\begin{equation}
H
=
-\frac12 \langle \Pi_B , \Pi_B \rangle
-\frac12 \langle H_B , H_B \rangle
+
G[Y,\phi,\Pi_\phi]
+
\langle H_A + \sigma , \Pi_A \rangle .
\label{H-canonical-general}
\end{equation}

The primary constraint \eqref{primary-general} must be preserved in time.  Since $\sigma$ appears in \eqref{H-canonical-general} only through the last term, one immediately finds
\begin{equation}
\dot \Pi_\sigma
=
- \frac{\delta H}{\delta \sigma}
=
- \Pi_A
\approx 0 .
\label{dot-Pisigma}
\end{equation}
Hence there is a secondary constraint
\begin{equation}
\Pi_A \approx 0 .
\label{secondary-general}
\end{equation}

The time evolution of $\Pi_A$ does not generate any further constraint.  Indeed, the only explicit $A$-dependence of \eqref{H-canonical-general} sits in
\begin{equation}
\langle H_A , \Pi_A \rangle
=
\langle \star_s \d A , \Pi_A \rangle .
\end{equation}
Its variation is proportional to $d(\star_s \Pi_A)$, and therefore vanishes once \eqref{secondary-general} is imposed.  Thus the Dirac procedure truncates.

Moreover,
\begin{equation}
\{\Pi_\sigma,\Pi_A\}=0 ,
\end{equation}
so $\Pi_\sigma$ and $\Pi_A$ are first class.  They are the Hamiltonian form of the $h$-gauge symmetry: the variables $A$ and $\Sigma$ are not independent physical fields, and only the gauge-invariant self-dual combination $F$ can carry physical information.

\subsection{Reduced Hamiltonian and Decoupling}

We now pass to the reduced phase space by imposing
\begin{equation}
\Pi_A = 0 .
\end{equation}
At this point it is natural to introduce the chiral combinations
\begin{equation}
\Pi_+ := \frac12 \bigl(\Pi_B + H_B\bigr),
\qquad
\Pi_- := \frac12 \bigl(\Pi_B - H_B\bigr).
\label{Pi-pm-def}
\end{equation}
Then
\begin{equation}
\Pi_B = \Pi_+ + \Pi_- ,
\qquad
H_B = \Pi_+ - \Pi_- ,
\qquad
\Pi_B - H_B = 2 \Pi_- .
\label{Pi-pm-rel}
\end{equation}
Using \eqref{H-canonical-general}, the reduced Hamiltonian becomes
\begin{equation}
H_{\rm red}
=
-\langle \Pi_+,\Pi_+\rangle
-\langle \Pi_-,\Pi_-\rangle
+
G[2\Pi_-,\phi,\Pi_\phi] .
\label{H-reduced-general}
\end{equation}
It is convenient to write this as
\begin{equation}
H_{\rm red}
=
H_{\rm extra} + H_{\rm phys},
\label{H-split-general}
\end{equation}
with
\begin{equation}
H_{\rm extra}
=
-\langle \Pi_+,\Pi_+\rangle ,
\label{H-extra-general}
\end{equation} 
\begin{equation}
H_{\rm phys}
=
-\langle \Pi_-,\Pi_-\rangle
+
G[2\Pi_-,\phi,\Pi_\phi] .
\label{H-phys-general}
\end{equation}
This is the desired decoupling.  The field $B$ contributes two chiral combinations, but only $\Pi_-$ appears in the interacting sector.  The other combination, $\Pi_+$, has become a completely free and completely decoupled extra sector.

To make this statement precise, let
\begin{equation}
I = [i_1 \cdots i_{2n}], \qquad
J = [j_1 \cdots j_{2n}]
\end{equation}
denote antisymmetric multi-indices.  The canonical Poisson brackets imply
\begin{equation}
\{\Pi_+^{\,I}(x),\Pi_+^{\,J}(y)\}
=
\frac12 \,
\epsilon^{I k J}\,
\partial_k \delta^{(4n+1)}(x-y),
\label{PB-plus}
\end{equation}
\begin{equation}
\{\Pi_-^{\,I}(x),\Pi_-^{\,J}(y)\}
=
-\frac12 \,
\epsilon^{I k J}\,
\partial_k \delta^{(4n+1)}(x-y),
\label{PB-minus}
\end{equation}
\begin{equation}
\{\Pi_+^{\,I}(x),\Pi_-^{\,J}(y)\}
= 0 .
\label{PB-mixed}
\end{equation}
Therefore the two sectors commute.  In particular, the free shadow sector generated by $\Pi_+$ is dynamically independent of the interacting sector generated by $\Pi_-$ and the matter fields.

\subsection{Identification of Physical Self-Dual Field}

On the reduced phase space, \eqref{Y-def} becomes
\begin{equation}
2\Pi_- = q + \frac{\delta L_i}{\delta q} .
\label{eq:q-Piminus-relation}
\end{equation}
Equivalently,
\begin{equation}
q
=
\frac{\delta G}{\delta Y}
\Biggr|_{Y=2\Pi_-} .
\label{q-from-G}
\end{equation}
Thus the full gauge-invariant field strength is reconstructed from $\Pi_-$ and the matter variables alone:
\begin{equation}
F
=
\d t \wedge q - \star_s q \quad \mathrm{with}
\quad
\star F = F .
\label{physical-F-general}
\end{equation}
The decoupled variable $\Pi_+$ does not appear in $F$, does not couple to the matter fields, and does not enter the physical Hamiltonian.

This proves that the interacting sector describes only self-dual degrees of freedom.  The pair $(A,\Sigma)$ does not introduce any extra propagating anti-self-dual mode; it merely provides a redundant but local description of the self-dual field, with the redundancy removed by the first-class constraints generated by $\Pi_\sigma$ and $\Pi_A$.

Specialising to the free theory where $L_i=0$
\begin{equation}
G[Y] = \frac12 \langle Y,Y\rangle .
\end{equation}
Then \eqref{H-reduced-general} becomes
\begin{equation}
H_{\rm red}^{\rm free}
=
-\langle \Pi_+,\Pi_+\rangle
+
\langle \Pi_-,\Pi_-\rangle .
\label{H-free-general}
\end{equation}
This makes the sign structure completely transparent.  With mostly plus signature, the extra sector has the wrong sign, exactly as expected from the sign of the kinetic term in \eqref{gen-action}, while the physical self-dual sector has positive energy.  In the free theory one also has
\begin{equation}
q = 2\Pi_- ,
\qquad
F = \d t \wedge (2\Pi_-) - \star_s (2\Pi_-) ,
\end{equation}
so the physical field is directly identified with the phase space co-ordinate $\Pi_-$.

\subsection{Summary: Hamiltonian Analysis} \label{sec:summ_H}

We summarise the findings of the previous sections here for convenience. For the new Lagrangian,
\begin{equation}
\mathcal L = \frac12\,\d B\wedge  \star \d B - \d B\wedge F + \mathcal L_i(F,\phi),
\qquad
F=\d A+ \star \d A-\Sigma,
\end{equation}
with $\Sigma$ self-dual and with $\mathcal L_i$ independent of the extra field $B$, the Hamiltonian analysis closely parallels Sen's construction.

The essential points are as follows. First, the canonical analysis produces two first-class constraints, reflecting the $h$-gauge symmetry. These constraints can be used to remove the gauge redundancy in $F$ by either gauge fixing $(\d+\star \d)A$ or $\Sigma$ to zero. Second, after passing to the reduced phase space and introducing the chiral variables $\Pi_\pm$, the Hamiltonian splits into two decoupled parts,
\begin{equation}
\mathcal H_{\mathrm{red}}
= -\Pi_+^2 + \Big[-\Pi_-^2 + \mathcal U(2\Pi_-,\phi,\pi_\phi,\phi')\Big].
\end{equation}
The variable $\Pi_+$ belongs to a free additional sector with degrees of freedom coming \textit{only} from the extra field $B$. It has negative energy because the $B$ kinetic term has the wrong sign in mostly plus signature. The variable $\Pi_-$, together with the matter fields in $\mathcal{U}$, carries the full interacting dynamics of the physical fields. Third, the physical sector is entirely encoded in the self-dual field strength
\begin{equation}
\qquad F=\star F,
\end{equation}
and $F$ is determined by $\Pi_-$ through the algebraic relation \eqref{eq:q-Piminus-relation} and the definition \eqref{eq:Fq-decomp}. Thus the interacting theory contains only self-dual degrees of freedom and there are no propagating \textit{anti-self-dual} parts in the physical sector.

The new action, therefore, has exactly the expected structure: a physical chiral boson interacting with other fields, plus an additional decoupled chiral boson with negative definite energy.


\section{Gauge Fixing The New Action}

So far the new action would seem to be a somewhat trivial extension of Sen's formalism, where we introduced additional pure gauge modes that are then removed by the $h$-gauge symmetry. However, we have an option to gauge fix $h$-gauge symmetry in multiple different ways. Being a gauge theory, different gauge fixed actions are dynamically equivalent.

Let us quickly review how one recovers Sen's action. Given the new action and $A$, we choose $h$ such that 
\begin{equation}
    (\d + \star \d) A^{\mathrm{new}} = \d A + \star \d A -\delta_h (\d A + \star \d A) = \d A + \star \d A - \Delta h = 0 \;.
\end{equation}
A straightforward analysis then tells us that we end up with Sen's action with the identification of $\Sigma$ with $- \mathcal{F}$ in \pref{eq:Sen}. Therefore, we can interpret Sen's action as a (partially) gauge fixed version of the new action. 

\subsection{Gauge Fixing to Potential Based Action}

Alternatively, we can choose $h$ such that we gauge fix $\Sigma$ to zero:
\begin{equation}
     \Sigma^{\mathrm{new}} = \Sigma -\delta_h \Sigma = \Sigma - \nabla h = 0 \;.
\end{equation}
The resulting action (let us momentarily only work with the free theory), after this gauge fixing, now reads:
 \begin{align} \label{eq:new_free_action}
     S[B,C] &= \frac{1}{2} \int \d B \wedge \star \, \d B - \int \d B \wedge F_A  \quad \mathrm{with} \quad F_A:= \d A + \star \d A \nonumber\\
     &= \frac{1}{2} \int \d B \wedge \star \, \d B - \int \d B \wedge \star \, \d A - \int \d B \wedge \d A \;.
 \end{align}
 Notice that the self-dual field strength $F$ is the desired physical degree of freedom now defined in terms of a $(2n)$-form $A$ as
\begin{equation}
    F_A = \d A + \star \d A \ .
\end{equation}
With a minor abuse of terminology, we will continue to refer to $A$ as the \textit{potential} or \textit{gauge field} and $F_A$ as its field strength throughout. It is crucial, however, to keep in mind that the above definition of the field strength is inherited from the field strength $F$ in the original action, which after gauge fixing is now reduced to $F_A$. 

There is a residual symmetry after gauge-fixing that leaves $F_A$ unchanged. Explicitly,
\begin{equation}
    \delta_\gamma A = \d^\dagger \gamma\, , \, \gamma = \star \gamma \,,\, \nabla \gamma = 0 \implies \delta_\gamma F_A = 0 \;.
\end{equation}
The transformation parameter $\gamma$ is a self-dual, harmonic $(2n+1)$-form. This is a higher-form global symmetry which we will call \textit{harmonic} symmetry. Note that this symmetry is the correct replacement of the familiar electric or magnetic higher-form symmetry for ordinary $p$-form electrodynamics, that selects either the curvature or its dual as a field-strength. In case of self-dual theories, our formalism shows that neither electric nor magnetic higher-form symmetry singles out the correct field-strength; this is achieved by the harmonic symmetry.
 
The last term in the action is a boundary term that will not affect the equations of motion. But its inclusion has some distinct advantages that will become apparent later.

Let us now turn the source back on. It is already at this point the advantages of this formalism over Sen's formalism will be apparent. The source term inherited from the parent action is 
     \begin{equation}
         \mathcal{L}_{\mathrm{int}} = F_A \wedge \Omega \;,
     \end{equation}
 where $\Omega$ is the anti-self-dual $(2n+1)$-form source. The relevant equations of motion now read
 \begin{equation}
     \d \star F_A = \d F_A = \star J \;,
 \end{equation}
once we identify $\d \Omega = \star J$, where $J$ is the more standard current $2n$-form. At the level of classical sources, this is sufficient. However, it is preferable to have a construction of the source in terms of local matter fields so that we can look into a combined interacting theory with dynamical matter. It was already pointed out in \citep{Aggarwal:2025fiq} that it is very difficult for $\Omega$ to be constructed using any matter fields which is local. 

On the other hand, the standard current $J$ always admits construction in terms of local matter fields. The disadvantage is that it couples directly to the potential $A$ and it sources both self-dual and the anti-self-dual part. However, this is not a problem in our formulation, since we have already established that the new action does not have any propagating anti-self-dual degrees of freedom. So we are at liberty to turn on a source term of the form
\begin{equation}
    \mathcal{L}_{\mathrm{int}} = A \wedge \star J \;,
\end{equation}
and it leads us to the desired equation of motion. Indeed, on-shell both interaction terms are identical once we identify $\d \Omega = \star J$. But the point is, even off-shell, we will replace the source $\Omega$ in terms of $J$. One can, at this point, freely turn on kinetic terms for the sources too.

We would like to emphasise that this is a perfectly consistent approach since even though there is a source for the anti-self-dual part, since the kinetic part of the action does not give it dynamics, it simply does not propagate and does not contribute to any computation of a physical observable. We will explicitly verify this fact for $D=4$ QEMD in \pref{sec:example}.

Note that, if we trade off the source $\Omega$ in favour of $J$, then the total action including the source term is no longer invariant under the harmonic symmetry. This is to be expected, since strictly speaking the role of the harmonic symmetry is to leave the self-dual sector unchanged and the source term with $J$, as discussed, also contributes to the anti-self-dual sector. This does not lead to any inconsistency, since recall, even in standard $p$-form electrodynamics, the electric or magnetic higher-form symmetry is a symmetry only for the free action and is broken by the source term. And as we discussed the sector that breaks this symmetry is completely inert, having no propagating degrees of freedom, so it has no bearing on the physical observables.

\subsection{Relation to Maxwell-like Action}
If we ignore the total derivative term in the new action, we notice that the extra field $B$ can be easily integrated out in the path integral after completing the square. To this end, let us start with the action sans the boundary term:
\begin{equation}
    S = \frac{1}{2} \int \d B \wedge \star \, \d B - \int \d B \wedge \star \, \d A - \int \mathcal{L}_{\mathrm{int}}(A,\phi) + \int \mathcal{L}_{\mathrm{kin}}(\phi) \; \;.
\end{equation}
The partition function is
\begin{equation}
    Z = \int \mathcal{D}B \,\mathcal{D}A\, \mathcal{D}\phi e ^{i S[B,A,\phi^I]} \;.
\end{equation}
Defining $\widetilde{B}:= B - A$, we see field redefinition has a trivial Jacobian, and the action now becomes
\begin{align}
    S[B,A\phi^I] &= S_1[\widetilde{B}] + S_2[A,\phi^I] \nonumber \\
    S_1[\widetilde{B}] &:=  \frac{1}{2}\int \d \widetilde{B} \wedge \star \, \d \widetilde{B} \\
    S_2[A,\phi] &:= - \frac{1}{2} \int \d A \wedge \star \, \d A - \int \mathcal{L}_{\mathrm{int}}(A,\phi) + \int \mathcal{L}_{\mathrm{kin}}(\phi)\;, \\
    \mathcal{D}B \,\mathcal{D}A\, \mathcal{D}\phi &= \mathcal{D}\widetilde{B} \,\mathcal{D}A\, \mathcal{D}\phi \;.
\end{align}
So we can integrate out the extra field (which now the path integral dictates must always be thought of as carried by the field $\widetilde{B}$) and express the physical partition function as
\begin{equation}
    Z_{\mathrm{phys}} = \int \mathcal{D}A\, \mathcal{D}\phi^I e^{i S_2[A,\phi^I]} \;.
\end{equation}

Something quite wonderful have happened here. First, unlike the Sen's formalism, the extra field decoupling can already be achieved in position space and therefore obviating any worries about introduction of this extra field $B$ into the original action. Here we also see the necessity of demanding the extra fields do not enter the interaction terms, otherwise we could not have integrated them out completely. In Sen's formalism to see such a decoupling one would need to go into momentum space and perform non-local field redefinitions or work at the level of Hamiltonian for a fully rigorous derivation. Second, the resulting action involving only the physical fields has just the Maxwell-like kinetic term for the gauge potential $A$. 

The reader might, at this stage, be concerned at this point that such an action is also written down in textbooks, but they famously fail to capture the self-duality which requires us to impose it by hand (pseudoaction formalism). However, note that in the pseudoaction, the field-strength is defined to be the curvature of the gauge field, i.e. $F_{\mathrm{pseudo}} := \d A$, which of course leads to vanishing of the standard kinetic term if one imposes self-duality on $F_{\mathrm{pseudo}}$. On the other hand, the new action defines the correct self-dual field-strength to be $F_A = \d A + \star \d A$, which still allows us to write the standard kinetic term down and give us the correct equations of motion without imposing any constraint by hand. 

We emphasise again, that the claim is not that the new formalism with the gauge fixing that sets $\Sigma$ to zero is same as Maxwell action. What we just demonstrated that the new action indeed leads to Maxwell action upon a specific gauge fixing but with the definition of the field strength to be uniquely defined to be $F_A$. Any other choice of field-strength at this point will be in tension with the gauge fixing condition for the $h$-gauge symmetry and be inconsistent. So the resultant Maxwell-like action, despite its superficial similarity, is conceptually radically different. The Maxwell-like action here is obtained after a gauge fixing of a parent action and that gauge fixing choice has several implications that the textbook pseudoaction do not possess. \\

In fact, at this point, we can distil a very pragmatic prescription. We can simply start with the Maxwell-like action (let us consider just the free theory since interaction has no bearing in the following)
\begin{equation}
    S_{\mathrm{Maxwell}} = \frac{1}{2} \int \d A \wedge \star \d A \; \overset{\text{EOM}}{\longrightarrow} \; \d \star \d A = 0 \;.
\end{equation}

The action has a $U(1)$ gauge symmetry $\delta A = \d \lambda$. But this alone does not specify that the field strength --- the physical observable in the theory --- has to be captured by the curvature of $A$. It is a prudent to ask: what tells us that the field strength should be the curvature? If we insist that the field strength be linear in first derivatives of $A$, the most general possibility allowed by gauge symmetry alone is 
\begin{equation}
    \mathscr{F} = a \, \d A + b \, \star \d A \;;\; a,b \in \mathbb{R} \text{ constants. }
\end{equation}

If we demand in addition a higher-form symmetry under which the field strength remains unchanged, then that singles out (up to an irrelevant overall normalisation) a unique combination of $a,b$. For example, demanding the electric higher-form symmetry $\delta A = \alpha$ for a closed $2n$-form $\alpha$ tells us the Bianchi identity must be $\d \mathscr{F} = 0$ and that tells us $b= 0$, $a=1$. On the other hand, demanding the magnetic higher-form symmetry tells us it must be $\star \mathscr{F}$ that satisfies the Bianchi, and singles out $a=0$ and $b=1$. Typically, one chooses one of these two over the other --- motivated principally by the kind of matter couplings in the theory. However, we see that neither of these two are the correct higher-form symmetry of the free action when it comes to a self-dual field-strength. The crucial observation is that the Bianchi identity can no longer be treated as an identity, both $\d \mathscr{F}$ and $\d \star \mathscr{F}$ must be equivalent, physically and mathematically. It cannot be that one satisfies an identity while the other satisfies a dynamical equation of motion. 

A simple investigation then identifies the correct higher-form symmetry to be demanded is precisely the one we observed before, viz. harmonic symmetry
\begin{equation} \label{eq:higher-form}
    \delta_\gamma A = \d^\dagger \gamma \;, \; \gamma = \star \gamma \;, \text{and}\; \Delta \gamma = 0\;
\end{equation}
 which uniquely singles out the correct field-strength to be the case $a=b=1$, i.e., 
 \begin{equation}
     F = \d A + \star \d A.
 \end{equation}

Invoking a different definition of field-strength other than the curvature of the gauge field is not new. Two noteworthy precursors, both for describing an action in presence of fundamental magnetic monopoles, are Dirac's original attempt \citep{Dirac:1948um} and Zwanziger's improvement of it in terms of two potentials \citep{Zwanziger:1970hk}. However, we wish to clarify that for our case this is not an ad hoc prescription, but it is derived by starting from an underlying gauge invariant action and by partially gauge fixing choosing a specific gauge choice.. 

Notice that the Maxwell-like action is invariant only up to a total derivative term under the harmonic symmetry \pref{eq:higher-form}. This is the price to pay for ignoring the total derivative term which was present in the new action in \pref{eq:new_free_action}. In the new action the higher-form symmetry is manifest. On the other hand, the Maxwell-like action is more familiar, free of any unphysical degrees of freedom, but the higher-form symmetry uniquely identifying the correct field-strength is not manifest.

At this point we would like to caution the reader once again that while working with Maxwell-like action seems to be the simplest option, the new formalism is not just the Maxwell-like action. Its parent action, which contains Sen's action as a gauge fixed version, only contains a self-dual field strength, and the Maxwell-like action is obtained by a partial gauge-fixing which, as a result, inherits the unique definition of the field strength which is also self-dual. It is no longer permitted to consider other field strengths which are invariant under $\mathrm{U}(1)$ gauge symmetry but not under the harmonic global symmetry. Other field-strengths, such as $\d A$ or $\d A - \star \d A$ would be inconsistent with the gauge fixing condition, which is now monitored by the harmonic global symmetry. 

Of course, it is always possible to gauge this harmonic global symmetry and return to the original gauge invariant form, which we now turn to discussing.

\subsection{Returning to The New Action}
Sen's formalism stems from the spacetime action for string theory \citep{Sen:2015nph,Mamade:2025jbs}. In particular, the RR-sector, even in the worldsheet theory, only enters via its field strength in the $(-1/2,-1/2)$ picture number. This suggest that the supergravity action obtained as a low-energy effective action of the string field theory action will treat all RR fluxes as dynamical variables without resorting to the potentials. This has indeed been verified recently in \citep{Mamade:2025jbs}. On the other hand, the textbook supergravity actions, at least for type IIA, seems to work flawlessly just with potentials. Of course, one can object this is not a fair comparison, since the SFT action once again treats both the electric and magnetic components of RR fluxes in equal footing whereas the type IIA action most definitely is written in terms of the electric variables. This seems to suggest that a true stringy action, in the canonical picture number, would necessarily involve working with only fluxes. The new action presented here, tells us there is a way to alleviate this tension. 

While the new action presented here allows us to work with potentials only after gauge fixing, it raises a crucial question: how exactly does a potential based description move over to a flux-based description? Understanding this transition is crucial if we are to understand the map between stringy representation of the supergravity fields and the standard field theory representation of the same fields. The relation turns out to be very simple, and natural. String theory does not possess any global symmetry \citep{Banks:1988yz,Banks:2010zn}. For the new formalism based on potential to therefore connect to stringy variables, we need to \textit{gauge} the harmonic symmetry.

This is done easily by introducing a background ($2n+1$)-form self-dual gauge field $\Sigma$ that transforms appropriately
\begin{equation}
    \delta_h A = \d^\dagger h \;, \delta_h \Sigma = \Delta h ,\; \Delta h \neq  0,
\end{equation}
and shift the field strength, redefining it as $ \mathcal{F} = F_A - \Sigma$ which is now gauge invariant, i.e. $\delta_h \mathcal{F} = 0$. Restoring the omitted boundary term now immediately leads us back to the new action.

 Note once again, this gauging is easily done in the new action with the extra fields. In the Maxwell-like theory this requires reinserting the omitted boundary term since the action cannot be written entirely using \textit{only} the correct field-strength.

All of the results obtained in Sen's formalism are readily subsumed in this new formalism. But this new formalism offers something extra: we can couple to matter via familiar minimal coupling prescriptions, and the decoupling of the shadow sector is very direct and simple. These advantages are more striking for the case of electromagnetic duality, to which we now turn. For the remainder of this draft, we will start directly with the partially gauge fixed form where $\Sigma$ (or its equivalent avatar) is gauge fixed to zero. The correct definition of the field-strength fixed by the appropriate harmonic higher-form symmetry. Gauging this harmonic global symmetry will always allow us to get back to the new action form if one so chooses. The new action then could always be gauge fixed to set the potential part to zero and give us back the Sen's action.

\section{The New Action: QEMD in \texorpdfstring{$D=4$}{D=4}}\label{sec:new_4D}

In \cite{Aggarwal:2025fiq}, following Sen's formalism \citep{Sen:2019qit}, we proposed a novel Lorentz- and
duality-invariant action for quantum electro-magnetodynamics
\emph{sans} gauge potentials and used to it to show that the it
reproduces the correct current correlation functions that can be
derived in an action-independent fashion
\cite{Weinberg:1965rz}. Notably, we used this Lagrangian to show that
the Dirac's charge quantisation condition is scale-invariant to all
orders in perturbation theory.

This action was formulated solely in terms of field strengths, which
in turn naturally couple to $ 2 $-form sources $ \Sigma_e $
and $ \Sigma_m $, which may be thought of as worldsheets whose
boundaries are the worldlines of electrically and magnetically charged
particles. The coupling counted the electromagnetic flux piercing the
worldsheets associated to charged particles. While our earlier results
did not require it, as pointed out in \pref{sec:intro}, a formalism in which
charged matter can be described in terms of the familiar $ 1 $-form
currents and kinetic terms is more advantageous. The goal of this section is to develop
precisely such a formalism, by leveraging the insights of the previous
section. 

\subsection{Action and the Equations of Motion}

The new action once again resembles the action written in \cite{Aggarwal:2025fiq} following Sen's formalism:
\begin{equation}
  \label{eq:old-action}
  \begin{aligned}
    S = \int _{\mathcal{M}} ^{} \bigg[& \frac{1}{2} \mathrm{d} B \wedge \star \mathrm{d} B + \frac{1}{2} \mathrm{d} \check{B} \wedge \star \mathrm{d} \check{B} - \mathrm{d} B \wedge \star F - \mathrm{d} \check{B} \wedge \star \widetilde{F}  \\
    &\quad - \frac{1}{2}  \widetilde{F}  \wedge \star \Sigma_m - \frac{1}{2} F  \wedge \star \Sigma_e  \bigg] \ .
  \end{aligned}
\end{equation}
However, unlike Sen's formalism, the field-strength is no longer treated as the fundamental dynamical variable. Instead following the motivations outlined in the previous section,
we \emph{define} the field strength to be
\begin{equation}
  \label{eq:field-strength-defn}
F = \mathrm{d} A + \star \mathrm{d} \check{A} \ , 
\end{equation}
which in turn implies the dual field-strength is
\begin{equation}
\widetilde{F} = \star \mathrm{d} A - \mathrm{d} \check{A}  \ .
\end{equation}

The field-strength is now defined in terms of a pair of potentials $A$ and $\check{A}$. Two-potential approach to QEMD is most certainly not new \citep{Zwanziger:1970hk}. However, our action has two crucial improvements over Zwanziger's approach --- our action is manifestly Lorentz invariant and our definition of field-strength in terms of the potentials is not \textit{ad hoc} but fixed by harmonic higher-form symmetry as we will elaborate shortly.

On plugging the expression of the field-strength into \cref{eq:old-action}, we see that the terms in
the Lagrangian that couple the additional fields to the
electromagnetic field strengths simplify as follows:
\begin{equation}
  \begin{aligned}
  - \mathrm{d} B  \wedge \star F &= - \mathrm{d} B  \wedge \star \mathrm{d} A + \mathrm{d} \left( B  \wedge \mathrm{d} \check{A}  \right) \ , \\
  - \mathrm{d} \check{B} \wedge \star \widetilde{F} &= + \mathrm{d} \check{B} \wedge \star \mathrm{d} \check{A} + \mathrm{d} \left( \check{B} \wedge \mathrm{d} A \right) \ .
  \end{aligned}
\end{equation}
Notice in particular the analogues of the boundary terms that also appeared in the previous section.

Further, the matter-photon interaction terms can also be simplified:
on defining
\begin{equation}\label{eq:current-2-form}
\begin{aligned}
\mathrm{d} ^{\dagger } \Sigma_m = j _{m} \quad \mathrm{and} \quad \mathrm{d} ^{\dagger } \Sigma_e = - j _{e} \ ,
\end{aligned}
\end{equation}
we can show that
\begin{equation}
- \frac{1}{2}  \widetilde{F}  \wedge \star \Sigma_m  - \frac{1}{2} F  \wedge \star \Sigma_e = A \wedge \star j _{e} + \check{A} \wedge \star j _{m} \ .
\end{equation}

On neglecting the boundary terms, we get the following  action
\begin{equation}
  \label{eq:sen-potential-action}
  \begin{aligned}
    S = \int _{\mathcal{M}} ^{} \bigg[& \frac{1}{2} \mathrm{d} B  \wedge \star \mathrm{d} B  + \frac{1}{2} \mathrm{d} \check{B} \wedge \star \mathrm{d} \check{B} - \mathrm{d} B  \wedge \star \mathrm{d} A + \mathrm{d} \check{B} \wedge \star \mathrm{d} \check{A} \\
    &\quad + A \wedge \star j _{e} + \check{A}  \wedge \star j _{m} \bigg] \ .
  \end{aligned}
\end{equation}
Here, $ B $ and $ \check{B} $ are additional $ 1 $-form fields,
$ A $ and $ \check{A} $ are $ 1 $-form gauge fields that together
constitute the photon field, and $ j _{e} $ and $ j _{m} $ are
electrically and magnetically charged $ 1 $-form sources.

\subsubsection*{Equations of Motion}\label{sec:EOM}
The equations of motion derived from this action are
\begin{equation}
  \label{eq:our-eoms}
\begin{aligned}
  \delta _{B}S = 0 &\Rightarrow \mathrm{d} \star \mathrm{d} B = \mathrm{d} \star \mathrm{d} A \ , \\
  \delta _{\check{B}}S = 0 &\Rightarrow \mathrm{d} \star \mathrm{d} \check{B} = -\mathrm{d} \star \mathrm{d} \check{A} \ , \\
  \delta _{A}S = 0 &\Rightarrow \mathrm{d} \star \mathrm{d} B = \star j _{e} \ , \\
  \delta _{\check{A} }S = 0 &\Rightarrow \mathrm{d} \star \mathrm{d} \check{B} = -\star j _{m} \ .
\end{aligned}
\end{equation}

On using \cref{eq:our-eoms} and the definition of the field strength
in \cref{eq:field-strength-defn}, we find that the equations of motion for the physical fields are
\begin{equation}
\begin{aligned}
  \mathrm{d} ^{\dagger }F &= j _{e} \ , \\
  \mathrm{d} F &= \star j _{m} \ .
\end{aligned}
\end{equation}
which are the generalised Maxwell equations in the presence of
electrically and magnetically charged matter. 

The additional fields $ B  $ and $ \check{B} $ must decouple and indeed they do. With the linear combinations
\begin{equation}
\begin{aligned}\label{H_fields}
  \mathcal{B} &=  B  -  A \ , \\
  \check{\mathcal{B}} &=  \check{B} +  \check{A} \ ,
\end{aligned}
\end{equation}
we can write rewrite the action \cref{eq:sen-potential-action} 
\begin{equation}
  S = \int _{\mathcal{M}} ^{} \left[\frac{1}{2} \d \mathcal{B} \wedge \star \d \mathcal{B} + \frac{1}{2} \d \check{\mathcal{B}} \wedge \star \d \check{\mathcal{B}} - \frac{1}{2} \mathrm{d} A \wedge \star \mathrm{d} A - \frac{1}{2} \mathrm{d} \check{A} \wedge \star \mathrm{d} \check{A} + A \wedge \star j _{e} + \check{A}  \wedge \star j _{m} \right] \ .
\end{equation}
Notice that the fields $ \mathcal{B} $ and $ \check{\mathcal{B}}$ are simply free
fields and completely decoupled, which we are free to integrate out to get
\begin{equation}
  \label{eq:doubled-maxwell}
  S = \int _{\mathcal{M}} ^{} \left[ - \frac{1}{2} \mathrm{d} A \wedge \star \mathrm{d} A - \frac{1}{2} \mathrm{d} \check{A} \wedge \star \mathrm{d} \check{A} + A \wedge \star j _{e} + \check{A}  \wedge \star j _{m} \right] \ .
\end{equation}
On doing so, the action looks like the Lagrangian for two copies of
Maxwell theory coupled to matter currents, as a consequence of ignoring
boundary terms in the passage from \cref{eq:old-action} to
\cref{eq:sen-potential-action}. This is perfectly fine in flat spacetime, but may have non-trivial consequences in manifolds with boundaries.

We caution the reader that this appearance does not imply we have a pair of Maxwell fields. The definition of the field-strength ensures that there is a single photon, captured by two potentials. But the photon can couple to both electric and magnetic sources while maintaining manifest duality. The invariance under the full duality rotation can be made manifest already at the level of action given in \cref{eq:sen-potential-action} using exactly the same duality doublets as defined in \citep{Aggarwal:2025fiq}.

We now discuss the (gauge) symmetries of the action.

\subsection{$U(1)$ Gauge Symmetry}

Let us start with the shadow sector $ 1 $-form fields $ B $ and
$ \check{B} $. Since these are additional fields which eventually
decouple from all other physical fields, we only discuss their gauge symmetries in the interest of completeness. 

The fields $ B  $ and $ \check{B} $ have usual $ U(1) $ gauge
symmetries
\begin{equation}
\delta B  = \mathrm{d} \phi _{1} \quad \mathrm{and} \quad \delta \check{B} = \mathrm{d} \phi _{2} \ ,
\end{equation}
for $ 0 $-forms $ \phi _{1} $ and $ \phi _{2} $. These gauge symmetries needs to be fixed in usual manner while deriving the propagator if one seeks to derive the propagator prior to performing the field-redefinition that decouples them. 

The $ 1 $-form gauge potentials $ A $ and $ \check{A} $ have similar
$ 0 $-form gauge symmetries
\begin{equation}
\delta A = \mathrm{d} \lambda \quad \mathrm{and} \quad \delta \check{A} = \mathrm{d} \check{\lambda } \ ,
\end{equation}
for $ 0 $-forms $ \lambda $ and $ \check{\lambda } $. These will also require being fixed prior to obtaining the propagator.

Since the new action
in \cref{eq:sen-potential-action} is formulated in terms of gauge
potentials, we can introduce matter couplings in the usual way
\begin{equation}
S \supset \int _{\mathcal{M}} ^{} \bigg[A \wedge \star j _{e} + \check{A} \wedge \star j _{m} \bigg] \ ,
\end{equation}
and the action is gauge invariant as long as the electric $ \left( j
  _{e} \right) $ and magnetic $ \left( j _{m} \right) $ currents are
separately conserved.

\subsection{One-Form Symmetries}
These extra fields possess $ 1 $-form symmetries
\begin{equation}
\delta B = \Phi  _{1} \quad \mathrm{and} \quad \delta \check{B} = \Phi  _{2} \ ,
\end{equation}
for a pair of closed $ 1 $-forms $ \Phi _{1} $ and $ \Phi _{2} $. 

It is not interesting to think of gauging these $1$-form symmetries, since due to decoupling they will only couple to a background gauge field which does not couple to the physical sector. So we will
not discuss these symmetries any further.

As is well-known, Maxwell theory without sources also has higher-form
symmetries. It is of particular interest to see how these are
manifested in our formalism. For most of this section, we will turn
off all source terms and consider free Maxwell theory. 

For closed $ 1 $-forms $ v $ and $ \check{v } $, the action in
\cref{eq:doubled-maxwell} is invariant under
\begin{equation}
  \begin{aligned}
    \text{electric $ 1 $-form symmetry:}& \quad \delta A = v  \ , \\
    \text{magnetic $ 1 $-form symmetry:}& \quad \delta \check{A} = \check{v } \ .
  \end{aligned}
\end{equation}
Notice that the field strength $ F $ defined in
\cref{eq:field-strength-defn} is invariant too. However, when working
with the more familiar looking action in \cref{eq:doubled-maxwell}, it
is not at all obvious that the electric and magnetic $ 1 $-form
symmetries have a mixed 't Hooft anomaly, i.e.~they cannot both be
gauged. An important demonstration of the consistency of our formalism
will be to show that this familiar result is reproduced. 

To prepare for a discussion of gauging these higher-form symmetries,
first, note that this $ 1 $-form symmetry can be uplifted to the
action in \cref{eq:sen-potential-action} straightforwardly. It will be
helpful here to restore the boundary terms that we neglected in the
passage from \cref{eq:old-action} to
\cref{eq:sen-potential-action}. Then the action we are considering is
then
\begin{equation}
  S = \int _{\mathcal{M}} ^{} \left[\frac{1}{2} \mathrm{d} B \wedge \star \mathrm{d} B + \frac{1}{2} \mathrm{d} \check{B} \wedge \star \mathrm{d} \check{B} - \mathrm{d} B \wedge \star F + \mathrm{d} \check{B} \wedge F \right] \ ,
\end{equation}
where $ F $ is given by \cref{eq:field-strength-defn}.  It is easy to
check that for closed $ 1 $-forms $ v $ and $ \check{v} $, the action
is invariant under
\begin{equation}
  \label{eq:em-1-form-symmetries}
\delta B  = \delta A = v \quad \mathrm{and} \quad \delta \check{B} = -\delta \check{A} = -\check{v} \ .
\end{equation}

Let us consider the task of gauging the electric $ 1 $-form symmetry
by choosing a $ 1 $-form gauge parameter $ v $ that is not closed,
i.e.~$ \mathrm{d} v \neq 0 $. In this case:
\begin{equation}
  \label{eq:electric-1-form-gauging}
\delta B = v \quad \mathrm{and} \quad \delta F = \mathrm{d} v \ .
\end{equation}
We add to the action the following terms:
\begin{equation}
S _{\mathrm{e}} = S + \int _{\mathcal{M}} ^{} b _{e} \wedge \star F - \frac{1}{2} b _{e} \wedge \star b _{e} \ ,
\end{equation}
where $ b _{e} $ is the background gauge field coupling to the
electric $ 1 $-form current $ F $. The term quadratic in $ b _{e} $ is
a local counterterm that ensures that the action is gauge invariant
under \cref{eq:electric-1-form-gauging} and
$ \delta b _{e} = \mathrm{d} v $ to up to boundary terms:
\begin{equation}
\delta S _{e} = \int _{\mathcal{M}} ^{} \mathrm{d} \left( \check{B} \wedge \mathrm{d} v \right) \ .
\end{equation}

We could just as well have gauged the magnetic $ 1 $-form symmetry by
considering a $ 1 $-form gauge parameter $ \check{v}  $ that is not
closed, i.e.~$ \mathrm{d} \check{v} \neq 0 $. Under the
transformations
\begin{equation}
\delta \check{B} = - \check{v} \quad \mathrm{and} \quad \delta F = \star \mathrm{d} \check{v} \quad \mathrm{and} \quad \delta b _{m} = - \mathrm{d} \check{v} \ ,
\end{equation}
the gauged action with appropriate local counterterm
\begin{equation}
S _{m} = S - \int _{\mathcal{M}} ^{} b _{m} \wedge F - \frac{1}{2} b _{m} \wedge \star b _{m} \ ,
\end{equation}
is invariant up to boundary terms:
\begin{equation}
\delta S _{m} = \int _{\mathcal{M}} ^{} \mathrm{d} \left( B \wedge \mathrm{d} \check{v}  \right) \ .
\end{equation}

Notice that if we gauge both electric and magnetic $ 1 $-form
symmetries, the action
\begin{equation}
S _{em} = S _{e} + \int _{\mathcal{M}} ^{} b _{m} \wedge F \ ,
\end{equation}
is no longer gauge invariant:
\begin{equation}
\delta S _{em} = \int _{\mathcal{M}} ^{} b _{m} \wedge \mathrm{d} v \ .
\end{equation}
This is precisely the (mixed) 't Hooft anomaly between electric and
magnetic $ 1 $-form symmetries. It is an obstruction to simultaneously
gauging global symmetries of the original theory, and represents a
non-trivial consistency check of our proposal.

\subsection{Harmonic Symmetry}

The symmetry that uniquely fixes the definition of the field-strength given in \pref{eq:field-strength-defn} is as follows:

\begin{equation}
    \delta A = \d^\dagger \Lambda \;, \; \delta \check{A} = \star \d \Lambda \;,
\end{equation}
for a $2$-form harmonic form $\Lambda$, i.e. $\Delta \Lambda = 0$. Under this transformation we have
\begin{equation} \label{eq:f-harmonic}
    \delta F = \Delta \Lambda = 0 \;.
\end{equation}

This is the manifestation of the harmonic $2$-form symmetry for QEMD in $D=4$ and we can see that the action in \pref{eq:sen-potential-action} is manifestly invariant whereas the pair of Maxwell action form is invariant only up to a total derivative.
We contend that for duality-invariant theories, this is the correct higher-form symmetry that should be imposed on the free theory.

Just like the self-dual case, this harmonic symmetry can be gauged. We introduce a background $2$-form gauge field $H$ such that the transformation now reads
\begin{equation}
    \delta A = \d^\dagger \Lambda \;, \; \delta \check{A} = \star \d \Lambda \;, \; \delta H = \Delta \Lambda \;, \; \text{with } \Delta \Lambda \neq 0 \;.
\end{equation}
Then the shifted field-strength $\mathcal{F} = F - H$ remains invariant, i.e. $\delta \mathcal{F} = 0$. Once again, if we now look at $\mathcal{F}$ closely
\begin{equation}
    \mathcal{F} = \d A + \star \d \check{A} - H \;,
\end{equation}
and we can then go back to the $h$-gauge invariant new action. Just like self-dual case, the $h$-gauge could be fixed to set $\d A + \star \d \check{A}$ to zero. The resulting action is precisely the Sen-like action introduced in \citep{Aggarwal:2025fiq}.
\subsection{Dirac Veto as $1$-form Gauge Symmetry}

Sen-like action for QEMD possesses an $1$-form gauge symmetry, even in presence of sources that is inherited by our action (with the gauge transformation now acting on the potentials instead of the field-strength). This reads
\begin{align}
    \delta B  &= \delta A = v \quad \mathrm{and} \quad \delta \check{B} = -\delta \check{A} = -\check{v} \nonumber \\
    \delta j _{e} &= \mathrm{d} ^{\dagger }\mathrm{d} v \quad \mathrm{and} \quad \delta j _{m} = \mathrm{d} ^{\dagger }\mathrm{d} \nonumber \check{v} \ ,
\end{align}
 with the $1$-`form parameters constrained to satisfy
 \begin{equation}
  \label{eq:Dirac-vetoes}
\int _{\mathcal{M}} ^{} v \wedge \star j _{e} = 0 \quad \mathrm{and} \quad \int _{\mathcal{M}} ^{} \check{v} \wedge \star j _{m} = 0 \ .
\end{equation}
Under these conditions, the action is invariant up to boundary terms as can be easily checked.

The conditions in \cref{eq:Dirac-vetoes} are precisely the Dirac veto
conditions, except we note that in our formalism these conditions
present themselves not as constraints on the worldsheets of Dirac
strings intersecting the worldlines of charged particles, but rather
as a requirement that the action is invariant under the aforementioned $1$-form gauge transformations. This is in synergy with the analysis presented in \citep{Hull:2024uwz} where it was shown that even for Dirac's original action, the Dirac veto can be interpreted as a higher-form symmetry. Therefore, it seems, this $1$-form gauge symmetry is more fundamental to QEMD, irrespective of whether it warrants an interpretation in terms of Dirac veto.

Note, the above transformation is for classical symmetry. For quantum theory, it suffices to relax the constraints on the $1$-form gauge parameters to 

\begin{equation}
  \label{eq:Dirac-vetoes_QM}
\int _{\mathcal{M}} ^{} v \wedge \star j _{e} = 2 \pi \mathbb{Z} \quad \mathrm{and} \quad \int _{\mathcal{M}} ^{} \check{v} \wedge \star j _{m} = 2 \pi \mathbb{Z}\ .
\end{equation}

\subsection{Charge Quantisation}

\begin{figure}[htbp] \label{fig:charge_quant}
  \centering
  \includegraphics[width=0.39\textwidth]{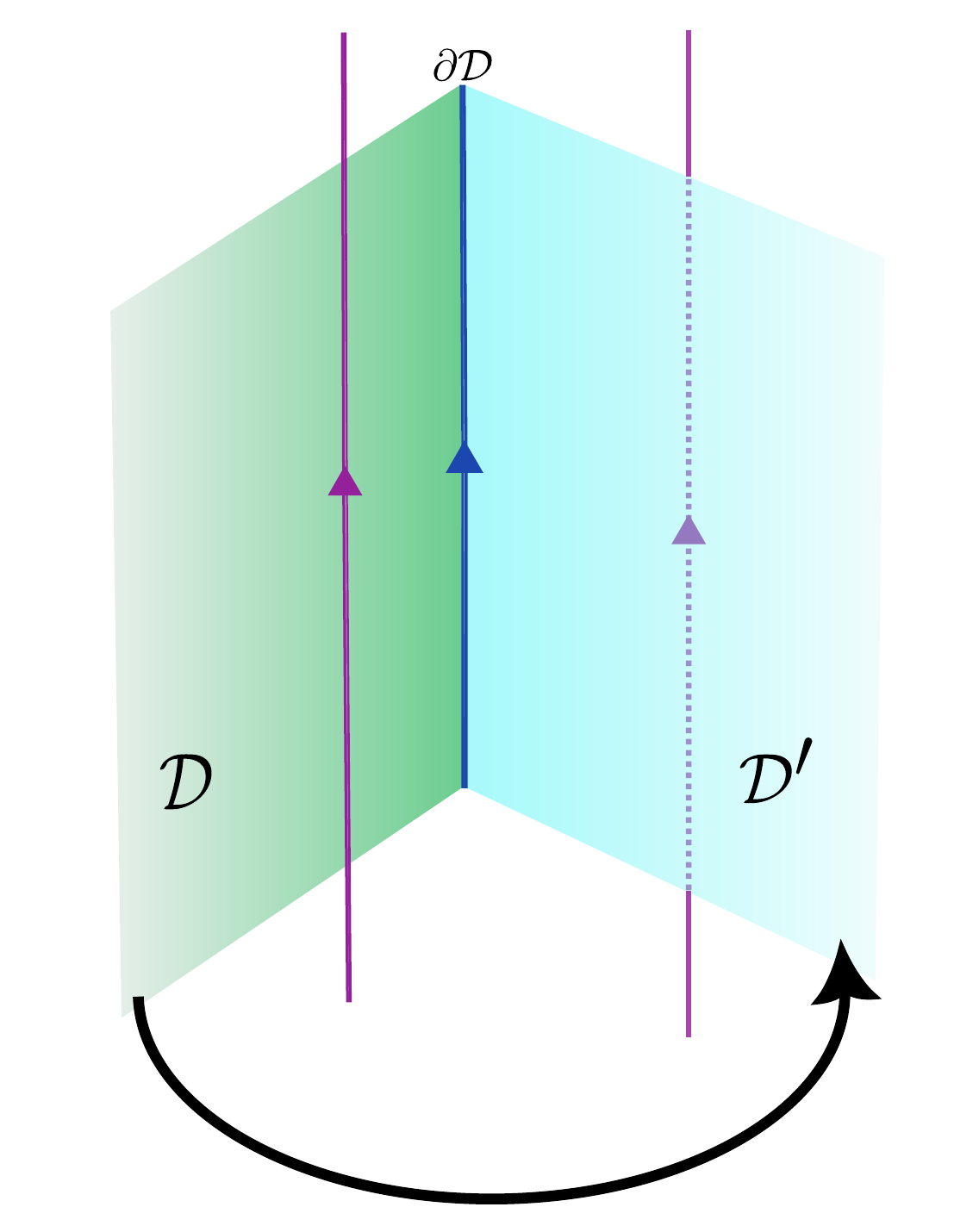}
  \caption{The surfaces $\mathcal{D}$ only has the constraint that it ends on the worldline of an electrically charged particle represented by the blue line. As it is deformed to a new surface $\mathcal{D}'$ while still attached to the same electric worldline, the action changes if there are magnetic worldlines represented by purple lines that are within the volume bounded by these two surfaces. Demanding the partition function remains unchanged leads directly to charge quantisation. }
\end{figure}

Another important consistency check for our formalism --- either in
its field strength or potential avatars --- is the demonstration that
our action is able to reproduce Dirac's charge quantisation
condition.

We'll use the following result: let $ \mathcal{D} $ be a
$ q $-submanifold of some $ n $-manifold $ \mathcal{M} $, with
boundary $ \partial \mathcal{D} $. Then for a
$ (q-1) $-form $ \sigma   $ we have Stokes' theorem:
\begin{equation}
\int _{\partial \mathcal{D}} ^{} \sigma   = \int _{\mathcal{D}} ^{} \mathrm{d} \sigma   \ ,
\end{equation}
We can also write this as
\begin{equation}
\int _{\mathcal{M}} ^{} \sigma  \wedge \star \delta _{\partial \mathcal{D}} = \int _{\mathcal{M}} ^{} \mathrm{d} \sigma \wedge \star \delta _{\mathcal{D}} \ ,
\end{equation}
where we have introduced $ \delta $-function-valued forms for later
convenience. Now we can write
\begin{equation}
\int _{\mathcal{M}} ^{} \sigma \wedge \star \delta _{\partial \mathcal{D}} = \int _{\mathcal{M}} ^{} \mathrm{d} \sigma \wedge \star \delta _{\mathcal{D}} = \int _{\mathcal{M}} ^{} \sigma \wedge \star \mathrm{d} ^{\dagger } \delta _{\mathcal{D}} \ ,
\end{equation}
and so we have the identity
\begin{equation}
  \label{eq:delta-adjoint-identity}
\delta _{\partial \mathcal{D}} = \mathrm{d} ^{\dagger }\delta _{\mathcal{D}} \ .
\end{equation}

Consider the familiar case of $ j _{e} $ corresponding to a particle
with charge $ -q $ moving along the worldline $ \mathcal{C} $. Then $
j _{e} = - q \delta _{\mathcal{C}} $. Since by
\pref{eq:current-2-form}, we have
\begin{equation}
\mathrm{d} ^{\dagger } \widetilde{\Sigma } = q \delta _{\mathcal{C}} \ ,
\end{equation}
then by \cref{eq:delta-adjoint-identity}, we have that
\begin{equation}
  \label{eq:sigma-for-worldline}
\widetilde{\Sigma } = q \delta _{\mathcal{D}} \ ,
\end{equation}
where $ \mathcal{D} $ is such that
$ \partial \mathcal{D} = \mathcal{C} $, i.e.~a surface whose boundary
is the worldline of the charged particle.

Now, the terms in the action with the source in
\cref{eq:sigma-for-worldline} can be written as
\begin{equation}
S \supset \int _{\mathcal{M}} ^{} F \wedge \star \widetilde{\Sigma } = q \int _{\mathcal{D}} ^{} F \ ,
\end{equation}
and since the surface $ \mathcal{D} $ is not uniquely specified by the
worldline, we must ensure that the path integral is single-valued
under a change in choice of surface.  So on considering how the action
in \cref{eq:old-action} changes when we change
$ \mathcal{D} \rightarrow \mathcal{D}' $ (see \cref{fig:charge_quant}), we get
\begin{equation}
  \begin{aligned}
    q \int _{\mathcal{D}'} ^{} F - q \int _{\mathcal{D}} ^{} F &= q \int _{\mathcal{D}'} ^{} F + q \int _{-\mathcal{D}} ^{} F \ , \\
    &= q \int _{\mathcal{S}} ^{} F \ ,
  \end{aligned}
\end{equation}
where $ \mathcal{S} $ is the $ 2 $-manifold that one gets from gluing
$ \mathcal{D}' $ to $ \mathcal{D} $, the latter taken with opposite
orientation. But $ \int _{\mathcal{S}} ^{} F $ is the total magnetic
charge contained inside $ \mathcal{S} $; let's say this is $ p $. The
path integral then acquires a phase $ e^{i q p}  $. For the theory to
be well-defined, this phase must be unity, so we must have
\begin{equation}\label{charge_quant}
q p = 2 \pi n \ ,
\end{equation}
for some $ n \in \mathbb{Z} $.

Note, this proof holds equally well if we treated $F$ itself as the dynamical variable sourced by the two form $\widetilde{\Sigma}$. Therefore, this proof also establishes the charge quantisation for Sen's formalism. While we only detailed the case for pure electric and magnetic charge, the methodology of the proof is similar for dyons. Indeed, the logical flow of the proof follows closely the original arguments due to Dirac \citep{Dirac:1948um} and the proof for dyons extends similarly for both Sen's and our formalism without any subtleties.

\subsection{The Witten effect}

We now demonstrate that in the presence of a CP-violating
$ \theta $-term, magnetic monopoles acquire an electric charge.

\begin{figure}
  \centering
  \includegraphics[width=0.39\textwidth]{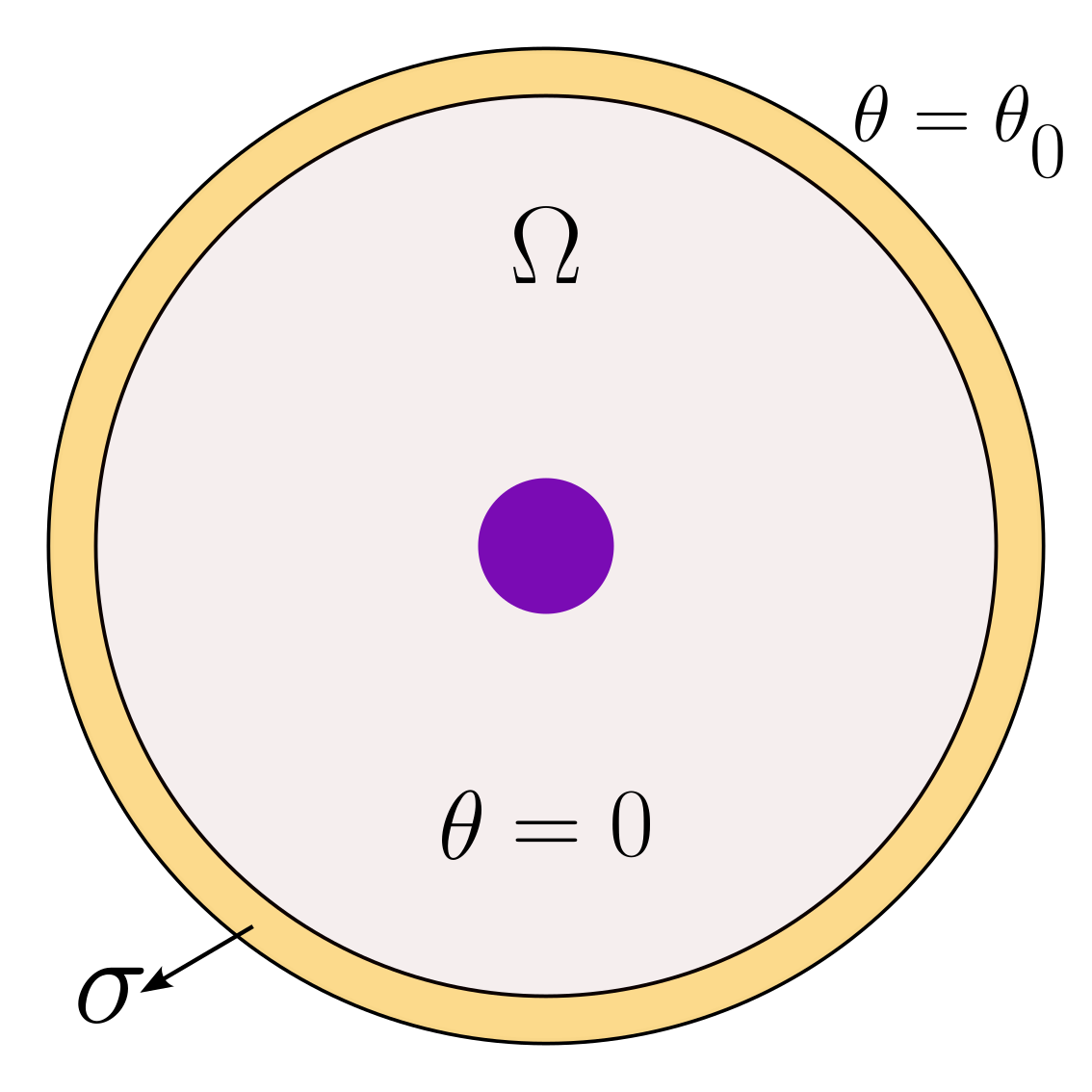}
  \caption{Monopole surrounded by axion domain wall. The theta vacua inside the shell is vanishing but outside the shell is a non-zero constant.}
  \label{fig:witten_eff}
\end{figure}

Consider the action $ S $ in \cref{eq:sen-potential-action} equipped with a theta term realised as a coupling to the axion field $\theta$:
\begin{equation}
S + \frac{1}{2} \int _{\mathcal{M}} ^{} \theta \, \mathrm{d} A \wedge \mathrm{d} A \ .
\end{equation}
 In the presence of the axion
field, the equations of motion are modified. Instead of simply finding
$ \mathrm{d} \star F = 0 $ where $ F $ is defined by
\cref{eq:field-strength-defn} in the absence of electric sources, we
find
\begin{equation}
  \mathrm{d} \star F =-\mathrm{d} \theta \wedge \mathrm{d} A
  \quad \Rightarrow \quad 
  \mathrm{d} \star \left( F - \theta \star  \mathrm{d} A \right) = 0 \ .
\end{equation}
To see the Witten effect, let us define
\begin{align}
    \mathcal{F}:=F - \theta \star  \mathrm{d} A\,,
\end{align}
and rewrite the equation of motion as
\begin{align}
    \d \star \mathcal{F}=0\,.
\end{align}
In the presence of the $\theta$ term, it is well known that the theory acquires an additional gauge symmetry reflecting the periodicity of $\theta$. This symmetry is simply the shift: $\theta \rightarrow \theta + 1$ \footnote{We work with field
  normalisations that set the radius of the axion to unity.}.
But under this transformation, $\mathcal{F}$ remains no longer invariant.
To remedy
this, we must simultaneously send
\begin{equation}
  \label{eq:Acheck-theta-transformations}
\theta \rightarrow \theta + 1 \quad \mathrm{and} \quad \check{A} \rightarrow \check{A} - A \ .
\end{equation}
This is precisely the Witten monodromy \cite{Witten:1979ey}, which implies—following the analysis of \cite{Wilczek:1987mv}—that monopoles acquire an electric charge. To see this explicitly, consider a monopole enclosed by an axion domain wall, as depicted in \cref{fig:witten_eff}.
 Then by Gauss's law the flux of the modified field strength $\mathcal{F}$ gives the shifted electric and magnetic charges in presence of a monopole
\begin{align}
    \int_{\sigma} \star \mathcal{F} &=\int_{\sigma} \star F +\int_{\sigma} \theta \d A\,,\cr
    \Rightarrow Q'_e &=Q_e -\theta_0 \Phi
\end{align}
where $\sigma$ is 2d the shaded region surrounding the monopole and its flux is given by\footnote{
The monopole flux can be shown to be $
   \theta_0\Phi \propto \int_{\Omega} \vec{B}\cdot \vec{\nabla}\theta $
where $\Omega$ is the three-dimensional volume bounded by the area $\sigma$ and $B_i=\epsilon_{ijk}\partial_j A_k$. This result was derived in \cite{Wilczek:1987mv}.
}
\begin{align}
\theta_0\Phi=\int_\sigma \theta \d A=-\int_\sigma A\wedge \d \theta \,.  
\end{align}
The usual story involving
monopoles being endowed with dyonic modes in axion electrodynamics, as
was recently reviewed in \cite{Heidenreich:2023pbi}, goes through
smoothly. Briefly, a heavy monopole can effectively be
described by an action $ S = \int _{\mathcal{C}} ^{} \check{A} $ along
some worldline $ \mathcal{C} $. In axion electrodynamics, requiring
the invariance under \cref{eq:Acheck-theta-transformations}
necessitates that we introduce a new compact degree of freedom
$ \sigma $ and an improved action
\begin{equation}
S _{\theta } = \int _{\mathcal{C}} ^{} \left[ \check{A} + \theta \left( \mathrm{d} \sigma + A \right) \right] \ ,
\end{equation}
together with the stipulation that under electric gauge
transformations $ A \rightarrow A + \mathrm{d} \lambda  $, the boson $
\sigma  $ shifts as $ \sigma \rightarrow \sigma - \lambda  $.  

Note that we could just as easily have added a ``non-standard axion
electrodynamics'' term of the form
\begin{equation}
S + \frac{1}{2} \int _{\mathcal{M}} ^{} \theta \,\mathrm{d} \check{A} \wedge \mathrm{d} \check{A} \ ,
\end{equation}
as discussed extensively in \cite{Heidenreich:2023pbi}. By
electric-magnetic duality, this induces a dual Witten effect that
endows electric charges with a magnetic charges under axion
shifts. We can repeat the flux calculation for $\widetilde{\mathcal{F}}$ and in that case we get the shift in the magnetic charge
\begin{align}
    Q'_m=Q_m-\theta_0\check{\Phi}\,,
\end{align}
where $\check{\Phi}$ is the dual monopole flux\footnote{It is given by $\theta_0 \check{\Phi}=\int_\sigma \check{\theta} \d \check{A}=\int_{\Omega} \vec{B}'\cdot \vec{\nabla}\check{\theta}$. where $B'_i=\epsilon_{ijk}\partial_j \check{A}_k$}.
As pointed out in \citep{Heidenreich:2023pbi}, it is perfectly consistent as a quantum field theory but are ruled out phenomenologically in four dimensions. In higher dimensions, however, no such phenomenological constraints are known to apply and such terms are perfectly sensible to consider.

It is also possible to turn on $\theta$ terms for both the gauge fields in this new formalism
\begin{align}
    S+\frac{1}{2}\int_{\mathcal{M}} \theta \d A\wedge \d A +\frac{1}{2}\int_{\mathcal{M}} \check{\theta} \d \check{A}\wedge \d \check{A}\,,
\end{align}
and without the electric and magnetic sources, the equations of motion are modified as follows
\begin{align}
     \d \star F=-\d \theta \wedge \d A \quad & \Rightarrow \quad \d \star (F-\theta \star \d A)=0 \,,\cr
     \d F =-\d \check{\theta} \wedge \d \check{A} \quad & \Rightarrow \quad \d \star (\widetilde{F}+ \check{\theta} \star \d \check{A})=0\,.
\end{align}
Again we define
\begin{align}
    \mathcal{F}:=F-\theta \star \d A\,,\quad \widetilde{{\mathcal{F}}}:=\widetilde{F}+ \check{\theta} \star \d \check{A}\,.
\end{align}
and rewrite the equations of motion as 
\begin{align}
    \d \star \mathcal{F}=0 \,,\qquad \d \star \widetilde{\mathcal{F}}=0\,.
\end{align}
As discussed earlier, this theory also has extra symmetries related to the periodicity in axion fields and to keep $\mathcal{F}$ and $\widetilde{\mathcal{F}}$ invariant, we must transform the gauge fields as follows
\begin{align}
   \theta \rightarrow \theta+1 \quad \Rightarrow \quad \check{A} \rightarrow  \check{A}+A\,,\cr
   \check{\theta} \rightarrow \check{\theta} +1 \quad \Rightarrow \quad A\rightarrow A-\check{A}\,.
\end{align}

Another natural topological boundary term in our formalism that can be considered is
\begin{align}\label{mixed_axion}
    S+ \int_{\mathcal{M}} \theta \,  \d A\wedge \d \check{A}\,,
\end{align}
but unlike the other two cases, it does \textit{not} produce a Witten effect. This can be seen as follows. The equations of motion without any electric and magnetic sources become
\begin{align}
    \d \star \mathcal{F}'=0\,,\qquad \mathcal{F}'= \d A -(\theta-1)\star \d \check{A}\,.
\end{align}
Now to make $\mathcal{F}'$ invariant under the $\theta$ gauge transformation, we must scale the gauge field as
\begin{align}
  \theta \rightarrow \theta +1 \quad \Rightarrow \quad  \check{A}\rightarrow 2\check{A}\,.
\end{align}
This scaling of the gauge field $\check{A}$ can be done by simply scaling the magnetic charge as $g\rightarrow \frac{g}{2}$.
Thus, this coupling in \eqref{mixed_axion} produces no shift in electric or magnetic charges and leads to no analogue of the Witten effect.

The presence of the gauge potentials therefore allows us to seamlessly incorporate Witten effect, which it is hard to envisage how to accommodate in Sen's formalism. Another obvious advantage of having the potential is the ability to couple to matter fields via time-tested prescription of minimal coupling. In \pref{sec:example} we write down explicit such matter coupling and exhaustively derive the Feynman rules to show how it subsumes, and indeed supersedes the analogous calculation performed in \citep{Aggarwal:2025fiq} using Sen's formalism.

\section{The New Action for EM duality in general dimensions}\label{sec:new_genD}
The new action is easily generalised to any dimension $d$ as follows. We introduce two gauge potentials, a $p$-form $A$ and a $(d-p-2)$-form $\check{A}$. The field-strength and its dual is then defined as

\begin{equation} \label{eq:correct_F-gen-d}
    F = \d A + \star \d \check{A} \;, \; \widetilde{F} = \star F = \star \d A + \eta \, \d \check{A} \;,
\end{equation}
 where we have written $\star^2 = \eta = - (-1)^{(p+1)(d-p-1)}$ which is same acting on a $(p+1)$-form and a $(d-p-1)$-form. Next we introduce the extra fields $B$ and $\check{B}$, with same rank as $A$ and $\check{A}$ respectively, to write the free action as

 \begin{equation}
     S_0 = \frac{1}{2}\int \d B \wedge \star \d B +  \frac{1}{2}\int \d \check{B} \wedge \star \d \check{B} - \int \d B \wedge \star F -\int \d \check{B} \wedge \star \widetilde{F} \;.
 \end{equation}
 Once again, there are two boundary terms, viz. $\d B \wedge \d \check{A} $ and $\d \check{B} \wedge \d A$ that do not play a role in obtaining the equation of motion, but are crucial in maintaining the manifest symmetry under the higher-form symmetry parametrised by a harmonic $(p+1)$-form $\Lambda^{(p+1)}$:
 \begin{equation} \label{eq:harmonic_gen-d}
     \delta A = \xi \,\d^\dagger \Lambda^{(p+1)}, \;\delta \check{A} = \star \d \Lambda^{(p+1)} \;, \text{with} \; \Delta \Lambda^{(p+1)} = 0 \implies \delta F = \delta \widetilde{F} = 0 \;.
 \end{equation}
Here the sign that accompanies the adjoint derivative is defined to be $\d^\dagger = \xi \, \star \d \star $ acting on a $p$-form. Explicitly,  $\xi= (-1)^{d(p+1)}$.

The coupling to source terms are easily introduced by considering a $p$-form current $J$ and a $(d-p-2)$-form current $\check{J}$ and writing the action
\begin{equation}
    S = S_0 + \int A \wedge \star J + \int \check{A} \wedge \star \check{J} \;.
\end{equation}

Dropping the boundary terms, and performing the following field redefinitions

\begin{equation}
    \mathcal{B} = B - A \; , \; \check{\mathcal{B}} = \check{B} - \eta \check{A} \;,
\end{equation}
 the action decouples into 
 \begin{equation}
     S = \frac{1}{2}\int \d \mathcal{B} \wedge \star \d \mathcal{B} +  \frac{1}{2}\int \d \check{\mathcal{B}} \wedge \star \d \check{\mathcal{B}} - \frac{1}{2} \int \d A \wedge \star \d A - \frac{1}{2} \int \d \check{A} \wedge \star \d \check{A} + \int A \wedge \star J + \int \check{A} \wedge \star \check{J} \;.
 \end{equation}
 At this point we can integrate out the extra fields and obtain the expected equations of motion for the physical sector
 \begin{align}
     \d \star \d A &= \star J \implies \d \star F = \star J \\
     \d \star \d \check{A} &= \star \check{J}  \implies \d F = \star \check{J} \;.
 \end{align}
 Once again, it suffices in flat space to work with a pair of Maxwell-like actions with usual coupling to source as long as one uses the correct field-strengths given in \pref{eq:correct_F-gen-d} as dictated by the invariance under the harmonic higher form symmetry defined in \pref{eq:harmonic_gen-d} instead of the curvatures $\d A$ and $\d \check{A}$. The Maxwell-like action is invariant only up to a boundary term under the harmonic higher-form symmetry. Just like the previous cases for $d=4$ and self-dual in $d=4n+2$, we will consider the action with the extra field to be primary knowing that for any calculations in flat space we can promptly integrate out the extra field to work with the familiar Maxwell-like action.

 Once again, this action has standard Abelian $p$-form gauge symmetries for the fields $A,B, \check{A}, $ and $\check{B}$. On top of that, they have the higher-form gauge symmetry parametrised by a $p$-form $v^{(p)}$ and a $(d-p-2)$-form $w^{(d-p-2)}$ as follows

 \begin{align}
     \delta A &= \delta B = v^{(p)} \;, \eta \, \delta \check{A} = \delta \check{B} = w^{(d-p-2)} \\
     \delta J &=(-1)^{p+1} \, \eta \, \xi \, \d^\dagger \d v^{(p)} \;, \; \delta \check{J} = (-1)^{d-p-1} \, \xi \, \d^\dagger \d w^{(d-p-2)}\,,
 \end{align}
where the gauge parameters satisfy 
\begin{align}
    &\int v^{(p)} \wedge \star J = 0 \; , \; \int w^{(d-p-2)} \wedge \star \check{J} = 0 \; \text{ (classical)} \\
    &\int v^{(p)} \wedge \star J = 2 \pi \mathbb{Z} \; , \; \int w^{(d-p-2)} \wedge \star \check{J} = 2 \pi \mathbb{Z} \; \text{ (quantum)} \;.
\end{align}
 These are higher-dimensional analogue of Dirac veto conditions in $d=4$ if one attempts to solve this system in terms of a purely electric (or magnetic) potential. The proof of charge (flux) quantisation for general $d$ proceeds exactly like the standard derivation. It is straightforward to check all the steps, for example following \citep{Deser:1997mz,Deser:1997se}, is easily adapted to this new action leading to an identical demonstration of charge (flux) quantisation of electric $(p-1)$-branes and magnetic $(d-p-3)$-branes. However, unlike the preceding approaches, we do not need to introduce any Dirac string-like objects since the two potentials allows us to completely evade any singularities in the space of the potentials, while at all times maintaining manifest duality and Lorentz symmetry.

 Notice that we can also easily realise the Witten effect (and its dual) for the charged branes in this duality invariant theory. However, unlike the $D=4$ case, there is no phenomenological input to set the dual Witten effects to zero and one can envisage a situation where both type of Witten effects are possibly present.

 Just like the self-dual and $d=4$ cases, the harmonic higher-form symmetry can be gauged. We introduce a $(p+1)$-form gauge field $H$ such that under the harmonic transformations
 \begin{equation}
     \delta A = \xi \,\d^\dagger \Lambda^{(p+1)}, \;\delta \check{A} = \star \d \Lambda^{(p+1)} \;, \delta H = \Delta \Lambda^{(p+1)} \;,
 \end{equation}
 allowing us to relax the condition $\Delta \Lambda^{(p+1)} = 0$ and obtain a gauge invariant field-strength
 \begin{equation}
     \mathcal{F} = F -H \;, \delta \mathcal{F} =  \delta (F - H) = \Delta \Lambda^{(p+1)} - \Delta \Lambda^{(p+1)} = 0 \;. 
 \end{equation}
 However, note that once again this field-strength is just offering a Hodge-decomposition of a $(p+1)$-form suggesting we simply adopt the action written in terms of the field-strength as the action but now the field-strength itself treated as a dynamical variable. This is precisely the Sen's formulation way of treating electromagnetic duality in general dimensions leading to the action that was first written down in \citep{Aggarwal:2025fiq}.

 \subsubsection*{Recovering Ordinary $p$-form Electrodynamics}

 It is quite easy to recover the standard $p$-form electrodynamics. We need to simply set the source for one of the gauge fields to zero. Let us take $\check{J}$ to be 0. Then the equations of motions reduce to 
 \begin{equation}
     \d \star F = \star J \;,\; \d F = 0 \;.
 \end{equation}

 In ordinary flat spacetime, we can always use Poincar\'e lemma to solve the latter only in terms of $A$ and effectively set $\check{A}=0$. This converts the latter equation of motion to the Bianchi identity, which is now true even \textit{off-shell}. However, note that with the definition of the field-strength with  $\check{A} \neq 0$ , the lack of magnetic charge is the same equation on-shell as Bianchi identity, but it is no longer obligated to remain true off-shell. In other words, we have managed to convert the Bianchi identity in an equivalent on-shell condition. This becomes advantageous in the self-dual case where the Bianchi identity is supposed to be \textit{only} on-shell statement. It is also advantageous in cases where the spacetime is not simply connected where the lemma only allows us to find such a solution locally. Typically, in such spaces one resorts to either using a gauge potential with singularity (eg. Dirac string) or adopt Wu-Yang construction \citep{Wu:1975es}. The advantage of a dual potential formalism is to avoid entirely the Wu-Yang construction of finding solutions patch by patch. Instead, relaxing the demand that field-strength be the curvature of a gauge potential we find that the field-strength defined in terms of dual potentials, which is fixed by demanding invariance under the harmonic higher-form symmetry, has regular solutions for both potentials, even in non-trivial spacetimes. We will elaborate on extension to non-trivial backgrounds in a follow-up article \citep{future}.

\section{Supersymmetry} \label{sec:susy}

The supersymmetrisation of the new action works very straightforwardly. Let us discuss the set-up for a Maxwell-like action of a $p$-form potential $A$ and its dual potential $(d-p-2)$-form $\check{A}$. The self-dual case is subsumed when we set $d= 4n+2$, $p = 2n$, and $A = \check{A} = C$.

So we have 
\begin{equation}
    S = -\frac1{2} \int \d A \wedge \star \d A - \frac{1}{2} \int \d \check{A} \wedge \star \d \check{A} \;.
\end{equation}
Suppose the supersymmetric completion of these action are as follows
\begin{equation}
    S_{_{SUSY}} = S + S_m[\phi^A] \;,
\end{equation}
where $S_m$ is a matter action involving rest of the matter fields (bosonic and fermionic) which we collectively denote as $\phi^A$, such that the entire supermultiplet is present and the action is supersymmetric under 
\begin{equation}
    \delta_S A = f[\phi^A] , \delta_S \check{A} = \check{f}[\phi^A], \delta_S \phi^A = G[A,\check{A}] \implies \delta_S S_{_{SUSY}} = 0 \;,
\end{equation}
for some known functions $f, \check{f},$ and $G$.
This supersymmetry can then be trivially extended to the action involving the extra field by simply demanding that the decoupled extra field do not transform under supersymmetry. Demanding $\mathcal{B} = B - A$ and $\mathcal{\check{B}} = \check{B} - \eta \check{A}$ are singlets fixes the supersymmetry transformation of $B$ and $\check{B}$ as 
\begin{equation}\label{susy_B}
    \delta_S B = \delta_S A = f[\phi^A] \;,\; \eta \, \delta_S \check{B} = \delta_S \check{A} = \check{f}[\phi^A] \;.
\end{equation}
With this it is now easily verified that the free action is supersymmetric even when written in terms of these extra fields. This supersymmetry will hold at the level of action and is essentially the same rules as the ones used for standard Maxwell-like actions. With the sources, of course, one needs to turn on appropriate interaction terms such that supersymmetry is maintained, just like in ordinary supersymmetric field theory. We will provide an explicit example of a supersymmetric action with interaction in \pref{sec:example}.

\section{Explicit example in \texorpdfstring{$D=4$}{D=4}} \label{sec:example}

The purpose of this section is to provide an explicit example of a QFT exhibiting the Abelian duality with explicit coupling to the matter sector. The $D=4$ is the best case to illustrate such a coupling. In so doing we will explicitly establish the equivalence of Feynman rules derived in Sen's formalism derived in \citep{Aggarwal:2025fiq} and in the new formalism. The new formalism, however, can work with explicit matter sector actions and therefore supersede the computations performed in \citep{Aggarwal:2025fiq}. While we explicitly work out a case in $D=4$, the arguments will hold for any general dimensions, including the self-dual cases. This section will establish using explicit tree-level and $1$-loop computations that turning on the standard current as sources do not affect the physical observables. We also present a brief summary of the supersymmetrisation of the QEMD action in new formalism following the general prescription outlined in \pref{sec:susy}.

\subsection{QEMD coupled to matter}

In this section, we shall consider matter coupled to the gauge fields $A$ and $\check{A}$ via the usual \textit{minimal coupling} prescription and compute various scattering amplitudes, in particular the 1-loop correction to either of the electric or magnetic charges. Since we are dealing with potentials now, we can couple the matter sector in the standard quantum field theoretic way.
We start with the action in \eqref{eq:sen-potential-action} in co-ordinate basis:
\begin{align}
    S_{A,\check{A}}=\frac{1}{2}\int d^4x \left[\partial_\mu B_\nu \partial^\mu B^\nu  +\partial_\mu \check{B}_\nu \partial^\mu \check{B}^\nu - \partial_\mu B_{\nu}\partial^\mu A^\nu +\partial_\mu \check{B}_{\nu}\partial^\mu \check{A}^\nu + j_{e,\mu}A^\mu + j_{m,\mu}\check{A}^\mu \right]\,.
\end{align}
For the matter sector, we consider magnetically charged complex scalar field and electrically charged Dirac spinor field. Thus the total action we consider is\footnote{The metric signature is $\lbrace -,+,+,+\rbrace$}
\begin{align}\label{action_total}
   S=\int d^4x \Big[\frac{1}{2}\partial_\mu B_\nu \partial^\mu B^\nu  +\frac{1}{2}\partial_\mu \check{B}_\nu \partial^\mu \check{B}^\nu-\frac{1}{2} \partial_\mu B_{\nu}\partial^\mu A^\nu +\frac{1}{2} \partial_\mu \check{B}_{\nu}\partial^\mu \check{A}^\nu - |\check{D}_\mu \phi|^2-i\bar{\psi}\slashed{D}\psi \cr
   -m_\phi^2 |\phi|^2-m_\psi \bar{\psi}\psi \Big]\,.
\end{align}
Here the gauge covariant derivatives $\check{D}$ and $D$ are defined w.r.t magnetic ($g$) and electric ($e$) couplings respectively:
\begin{align}
    \check{D}_\mu:=\partial_\mu + ig \check{A}_\mu\,, \qquad \slashed{D}:=\slashed{\partial}+i e \slashed{A}\,,
\end{align}
and the minimal interaction terms $j_e\cdot A$ and $j_m\cdot \check{A}$ are already contained in the gauge invariant matter kinetic terms:
\begin{align}
    j_m\cdot \check{A} &=ig\phi^\dagger \overleftrightarrow{\partial}_\mu \phi \check{A}^\mu \,,\cr
    j_e\cdot A &= e\bar{\psi}\gamma_\mu \psi A^\mu\,.
\end{align}
 As discussed earlier,
the electric and magnetic couplings are not independent; they are related through the Dirac quantisation condition \eqref{charge_quant}. This relation implies that both couplings cannot be treated perturbatively at the same time—only one can be treated in a perturbative expansion while the other remains non-perturbative. In the following analysis, we will treat the electric coupling $e$ as the perturbative parameter and compute its \textit{perturbative} one-loop correction.

\subsubsection{Feynman rules}

Before delve into the calculation of charge renormalisation, let us spell out the necessary Feynman rules. To extract the Feynman rules, we first gauge fix the action in \eqref{action_total} by introducing Lorentz gauge fixing terms for all the gauge fields in the action
\begin{align}
    S_{GF}= \frac{1}{2}\int d^4x \left[\left(\partial_\mu B^{\mu}\right)^2 +\left(\partial_\mu \check{B}^{\mu}\right)^2 +\left(\partial_\mu A^{\mu}\right)^2 +\left(\partial_\mu \check{A}^{\mu}\right)^2 \right]\,.
\end{align}
Let us start by finding the propagators rules for the gauge fields $A^\mu$ and $\check{A}^\mu$. For this purpose, we can work with the following part of the full action written in momentum space\footnote{
We use the following Fourier transform to express the position space fields in momentum basis
\begin{align}
    \Phi(x)= \int \frac{d^4k}{(2\pi)^4}e^{ik\cdot x} \Phi(k)\,.
\end{align}
}
\begin{align}
    S\supset \frac{1}{2}\int  \frac{d^4k}{(2\pi)^4} k^2 \left[ B_\mu(k) B^{\mu}(-k)+ \check{B}_\mu(k) \check{B}^{\mu}(-k) - B_{\mu}(k) A^\mu(-k) + \check{B}_{\mu}(k) \check{A}^\mu(-k) \right]\,.
\end{align}
It was shown in earlier in section \ref{sec:EOM} that although the physical fields: $(A^\mu,\check{A}^\mu)$ and extra fields: $B^{(i),\mu}$ mixes kinematically in position space, they decouples completely after a simple field redefinition. We can show this in momentum space as well. This also allows us to extract the Feynman rules cleanly. Let us define the following fields:
\begin{align}
    \widehat{B}_\mu^{(1)}:=B_\mu -\frac{1}{2} A_\mu\,, \qquad \widehat{B}_\mu^{(2)}:=\check{B}_\mu +\frac{1}{2}\check{A}_\mu\,.
\end{align}
In terms of the new fields $\widehat{B}^{(i)}_\mu$ and physical fields, we can rewrite the above action as follows
\begin{align}
    S \supset \frac{1}{2}\int  \frac{d^4k}{(2\pi)^4} k^2 \left[\sum_{i=1,2} \widehat{B}^{(i)}_\mu(k) \widehat{B}^{(i)\mu}(-k)-\frac{1}{4}A^\mu(k)A_\mu(-k) -\frac{1}{4}\check{A}^\mu(k) \check{A}_\mu(-k) \right]\,.
\end{align}
We see that the kinetic terms in the action has now splits into two part: a) kinetic term for the additional fields $\widehat{B}^{(i)}_\mu$ and b) kinetic term for the photon fields $A^\mu$ and $\check{A}^\mu$. From this action, we can write the photon propagators in the Lorentz gauge straightforwardly 
\begin{align}\label{feynman_photon}
    \Delta_{A-A}^{\mu \nu}(k) =\frac{1}{8}\frac{\eta^{\mu \nu}}{k^2+i\epsilon}\,,\qquad
    \Delta_{\check{A}-\check{A}}^{\mu \nu}(k) =\frac{1}{8}\frac{\eta^{\mu \nu}}{k^2+i\epsilon}\,.
\end{align}
The two point functions between the photon fields can also be found as
\begin{align}\label{A-A_function}
    \langle A^\mu(k)A^\nu(k')\rangle &=\frac{i}{8}\frac{\eta^{\mu \nu}}{k^2+i\epsilon}\delta^{(4)}(k+k')\,,\cr
    \langle \check{A}^\mu(k) \check{A}^\nu(k')\rangle &=\frac{i}{8}\frac{\eta^{\mu \nu}}{k^2+i\epsilon}\delta^{(4)}(k+k')\,,\cr
    \langle \check{A}^\mu(k) {A}^\nu(k')\rangle &=0\,.
\end{align}
Given the standard matter couplings in the action in \eqref{action_total}, we can easily write down Feynman rules involving various fields
\begin{align}\label{feynman_matter}
    \Delta_{\phi-\phi^\dagger}(p) &=\frac{-i}{p^2+m_\phi^2+i\epsilon}\,, \cr
    \Delta_{\psi-\bar{\psi}}(p) &=\frac{-i(\slashed{p}-m_\psi)}{p^2+m_\psi^2+i\epsilon}\,,\cr
    V^\mu_{\phi}(p_1,p_2,k) &=-g(p_1+p_2)^{\mu}\,, \cr
    V^\mu_{\psi}(p_1,p_2,k) &=ie\gamma^{\mu}\,,\cr
    V^{\mu \nu}_{\phi-A}&=2ig^2 \eta^{\mu \nu}\,.
\end{align}
Here we denote the four momenta of massive particles as $p_{1,2}^\mu$ with in-out going convention and massless momentum as $k^\mu$. 

\subsubsection{Scattering amplitudes: Charge Renormalisation}

\begin{figure}
	\centering
	\begin{tikzpicture}[scale=0.75]
\begin{feynman}
\vertex (i1) at (-2,1.5) {$p_2$};
\vertex (i2) at (-2,-1.5) {$p_1$};
\vertex (v1) at (0,0);

\vertex (v2) at (2.8,0);
\vertex (o1) at (5,1.5) {$p_3$};
\vertex (o2) at (5,-1.5) {$p_4$};

\vertex (ii1) at (6,1.5) {$p_3$};
\vertex (ii2) at (6,-1.5) {$p_1$};
\vertex (vv1) at (8.1,0);

\vertex (vv2) at (10.8,0);
\vertex (oo1) at (13,1.5) {$p_2$};
\vertex (oo2) at (13,-1.5) {$p_4$};

\diagram*{
 (i1) -- [anti fermion] (v1) -- [photon, edge label'=$k^\mu$] (v2),
 (i2) -- [fermion] (v1),
 (v2) -- [fermion] (o1),
 (v2) -- [anti fermion] (o2),
};
\diagram*{
 (ii1) -- [anti fermion] (vv1) -- [photon, edge label'=$k^\mu$] (vv2),
 (ii2) -- [fermion] (vv1),
 (vv2) -- [fermion] (oo1),
 (vv2) -- [anti fermion] (oo2),
};


\node at (1.0,-1.0) {$s$-channel};
\node at (9.6,-1.0) {$u$-channel};
\end{feynman}
\end{tikzpicture}
	\caption{$s$ and $u$ channel diagrams at tree level}
	\label{fig:tree}
\end{figure}
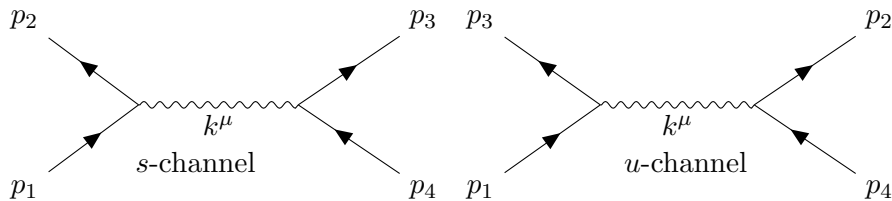

We can use the Feynman rules given in \eqref{feynman_photon} and \eqref{feynman_matter} to compute scattering amplitudes. As discussed earlier, we treat the electric charge $e$ as the perturbative coupling. Therefore, it is convenient to consider amplitudes that admit a Feynman diagram expansion order by order in $e$. In this section, we compute the electron-electron scattering amplitudes in this theory using standard perturbative methods. In particular, we compute the amplitudes related to the electric charge renormalisation up to 1-loop. We shall reproduce the calculations of \cite{Aggarwal:2025fiq} and show RG invariance of charge quantisation but with traditional matter currents defined earlier. This is one of the prominent advantage of this new potential based formalism.

Let us consider the amplitude of a scattering process for the following in-out going momentum configuration:
\begin{align}
    e^-_1+e^-_4 \rightarrow e^-_2+e^-_3 \,.
\end{align}
At tree-level, there are two possible Feynman diagrams: $s$ and $u$-channel, depicted in Fig. \ref{fig:tree}. We define the Mandelstam variables as: $s=(p_1-p_2)^2, u=(p_1-p_3)^2$ and $t=(p_1+p_4)^2$ with $p_i^2=-m_\psi^2$.
The contributions from individual diagrams can be easily computed and we get the tree-level amplitude as\footnote{For external fermion lines, we use $u$ spinors to denote electrons and $v$ spinors to denote positrons.}
\begin{align}\label{tree-level}
    \mathcal{A}^{(0)}_4 \left[p_1,p_4\rightarrow p_2,p_3\right] &=-\frac{e^2}{8}\left[\frac{\bar{u}_2(p_2)\gamma^\mu u_1(p_1)\bar{u}_3(p_3)\gamma_\mu u_4(p_4)}{s}-\frac{\bar{u}_3(p_3)\gamma^\mu u_1(p_1)\bar{u}_2(p_2)\gamma_\mu u_4(p_4)}{u} \right]\,,\cr
\end{align}
where the relative sign is due to the exchange of two fermions.

\begin{figure}
	\centering
	\includegraphics[width=0.57\textwidth]{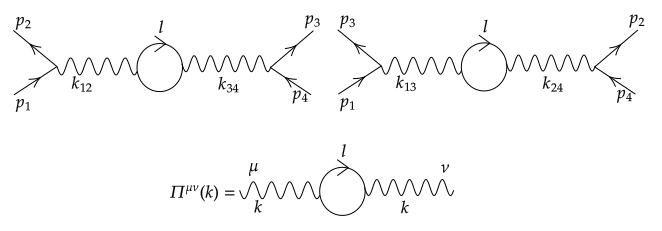}
	\caption{1-loop diagrams contributing to charge renormalisation.}
	\label{fig:qed_1_loop}
\end{figure}

At 1-loop there are two diagrams in Fig. \ref{fig:qed_1_loop} contributing to the vacuum polarisation ($\Pi^{\mu \nu}(k)$) related to electric charge renormalisation. Both the diagrams again can be computed straightforwardly and we write it as
\begin{align}
    \mathcal{A}_4^{(1)} \supset -e^2\bar{u}_2(p_2)\gamma^\mu u_1(p_1)\frac{\Pi_{\mu \nu}(k_{12})}{64s^2}\bar{u}_3(p_3)\gamma^\nu u_4(p_4) -(p_2\leftrightarrow p_3)\,,
\end{align}
Here $k_{12}=(p_1-p_2)=(p_3-p_4)$, due to momentum conservation. The 1-loop correction to the electric charge is captured by $\Pi^{\mu \nu}(k_{12})$ given by the following 1-loop integral 
\begin{align}
    \Pi^{\mu \nu}(k_{12}) &=(ie)^2\int \frac{d^4k}{(2\pi)^4}\frac{\text{Tr}\left[ \gamma^\mu (\slashed{l}-m_\psi)\gamma^\nu (\slashed{l}-\slashed{k}_{12}-m_\psi)\right]}{(l^2+m_\psi^2)[(l-k_{12})^2+m_\psi^2]}\,.
\end{align}
The evaluation of this loop integral using the Feynman parametrisation is straightforward and we skip the details and give the result below
\begin{align}
    \Pi^{\mu \nu}(k)= e^2(k^\mu k^\nu-k^2\eta^{\mu \nu})\Pi(k^2)\,,
\end{align}
where\footnote{Since $e$ has mass dimension $[e]_m=\frac{4-d}{2}$, we rescale: $e\rightarrow \mu^{\frac{4-d}{2}}e$ such that $e$ becomes dimensionless and $[\mu]_m=1$.} 
\begin{align}
    \Pi(k^2):=\Gamma(2-d/2)\mu^{4-d}\int_0^1 dx\, x(1-x)\left(m_\psi^2-k^2x(1-x)\right)^{\frac{d}{2}-2}\,.
\end{align}
From the structure of $\Pi^{\mu \nu}(k)$ it is clear that the Ward identity is satisfied
\begin{align}\label{pi_tensor}
 \Pi^{\mu \nu}(k)k_\mu=0= \Pi^{\mu \nu}(k)k_\nu\,.
\end{align}
Including the tree-level amplitude in \eqref{tree-level}, we obtain the 1-loop contribution to charge renormalisation 
\begin{align}
    \mathcal{A}_4^{(0+1)}\supset -\frac{e^2}{8}\Big[\bar{u}_2(p_2)\gamma_\mu u_1(p_1)\left( \frac{\eta^{\mu \nu}}{s}+\frac{\Pi^{\mu \nu}(k_{12})}{8s^2}\right)\bar{u}_3(p_3)\gamma_\nu u_4(p_4)\cr
    \qquad -\bar{u}_3(p_3)\gamma_\mu u_1(p_1)\left( \frac{\eta_{\mu \nu}}{u}+\frac{\Pi_{\mu \nu}(k_{13})}{8u^2}\right)\bar{u}_2(p_2)\gamma_\nu u_4(p_4) \Big]\,.
\end{align}
Using the E.O.M for the spinor wavefunctions $u,v$, we can simplify the above and write\footnote{Particularly we used the following identity follows from the E.o.M:
\begin{align}
 \bar{u}_2(p_2)\gamma_\mu u_1(p_1)(p_1-p_2)^\mu=0\,.   
\end{align}
}
\begin{align}
    \mathcal{A}_4^{(0+1)}\supset -\frac{e^2}{8}\Big[\frac{1}{s}\bar{u}_2(p_2)\gamma_\mu u_1(p_1)\left( 1+\frac{\Pi(s)}{8}\right)\bar{u}_3(p_3)\gamma^\mu u_4(p_4)\cr
    \qquad -\frac{1}{u}\bar{u}_3(p_3)\gamma_\mu u_1(p_1)\left(1+\frac{\Pi(u)}{8}\right)\bar{u}_2(p_2)\gamma^\mu u_4(p_4) \Big]\,.
\end{align}
Let us define the renormalized electric charge at some scale $\Lambda$ as
\begin{align}
    e^2_{R}(\Lambda)=e^2\left(1+\frac{\Pi (\Lambda)}{8} \right)\,,
\end{align}
then we can rewrite the amplitude up to 1-loop as
\begin{align}
    \mathcal{A}_4^{(0+1)}\supset -\frac{e^2_{R}(\Lambda)}{8}\Bigg[\frac{1}{s}\bar{u}_2(p_2)\gamma_\mu u_1(p_1)\left( 1+\frac{\Pi(s)-\Pi (\Lambda)}{8}\right)\bar{u}_3(p_3)\gamma^\mu u_4(p_4)\cr
    \qquad -\frac{1}{u}\bar{u}_3(p_3)\gamma_\mu u_1(p_1)\left(1+\frac{\Pi(u)-\Pi (\Lambda)}{8}\right)\bar{u}_2(p_2)\gamma^\mu u_4(p_4) \Bigg]\,,
\end{align}
where $(\Pi(s)-\Pi (\Lambda))$ and $(\Pi(u)-\Pi (\Lambda))$ are finite form-factor corrections. The analysis can be repeated for the magnetic charge by treating it as the perturbative coupling, in which case the relevant amplitudes would involve scattering processes of complex scalar states.

Now, we can follow the arguments presented in \cite{Aggarwal:2025fiq} to show that the charge quantisation condition \eqref{charge_quant} remains unchanged under renormalisation
\begin{align}
    e_R (\Lambda) g_R(\Lambda)=e g\,.
\end{align}
We have demonstrated that perturbative amplitudes in this theory \eqref{action_total} can be computed using standard quantum field–theoretic techniques. As an illustration, we present the analysis of the one-loop correction to the electric charge, where the renormalisation proceeds exactly as in QED. 
It is also worth emphasizing that the new potential-based formalism improves upon the earlier field-strength formulation of \cite{Aggarwal:2025fiq}. In particular, quantities such as $\Pi(k^2)$ can now be computed explicitly once the matter currents are specified, making the perturbative structure more transparent and fully accessible. Any other scattering process in this new formalism can be treated just as straightforwardly.

\subsubsection{Field strength based perturbation theory}

In this section, we shall connect the new potential formalism for the quantum magnetic-electrodynamics to the field strength based formalism in \cite{Aggarwal:2025fiq}. These two formalisms are result of two different gauge fixing choice of $h$-gauge symmetry of the parent action. So it stands to reason they will be completely equivalent. We will provide the explicit equivalence of the two set of Feynman rules here as a consistency check of that fact.

To start with, let us compute the two point function $\langle F_{\mu \nu}(k)F^{\rho \sigma}(k')\rangle$ in this new formalism.
The coordinate and momentum space representation directly follows from the definition \eqref{eq:field-strength-defn}
\begin{align}
    F_{\mu \nu}(x) &=\partial_{[\mu}A_{\nu]}(x)+ \epsilon_{\mu \nu \rho \sigma} \partial^\rho \check{A}^\sigma(x)\,,\cr
    F_{\mu \nu} (k)&=i\left(k_{[\mu}A_{\nu]}(k)+\epsilon_{\mu \nu \rho \sigma}k^\rho \check{A}^\sigma(k) \right)\,.
\end{align}
The calculation of the two-point function using \eqref{A-A_function} is straightforward and we get
\begin{align}
    \langle F_{\mu \nu}(k)F^{\rho \sigma}(k')\rangle &=-\left\langle \left(k_{[\mu}A_{\nu]}(k)+\epsilon_{\mu \nu \alpha \beta}k^\alpha \check{A}^\beta (k) \right) \left(k'^{[\rho}A^{\sigma]}(k')+\epsilon^{\rho \sigma \gamma \delta}k'_\gamma \check{A}_\delta (k')\right) \right\rangle \cr
    &=-k_{[\mu}k'^{[\rho} \langle A_{\nu]}(k)A^{\sigma]}(k')\rangle -\epsilon_{\mu \nu \alpha \beta}\epsilon^{\rho \sigma \gamma \delta}k^\alpha k'_\gamma \langle \check{A}^\beta (k)\check{A}_\delta (k')\rangle \cr
    &=\frac{i}{8}\frac{1}{k^2}\delta^{(4)}(k+k')\left[k_{[\mu}k^{[\rho} \delta_{\nu]}{}^{\sigma]}+\epsilon_{\mu \nu \alpha \beta}\epsilon^{\rho \sigma \gamma \beta}k^\alpha k_\gamma \right]\cr
    &=\frac{i}{4}\delta^{(4)}(k+k')\left[\frac{1}{k^2} k_{[\mu}k^{[\rho} \delta_{\nu]}{}^{\sigma]} -\frac{1}{4}\delta_{[\mu}{}^{[\rho}\delta_{\nu]}{}^{\sigma]} \right]\,.
\end{align}
In the last line, we used the following identity
\begin{align}
    \epsilon_{\mu \nu \alpha \beta}\epsilon^{\rho \sigma \gamma \beta}k^\alpha k_\gamma =k_{[\mu}k^{[\rho} \delta_{\nu]}{}^{\sigma]}-\frac{k^2}{2}\delta_{[\mu}{}^{[\rho}\delta_{\nu]}{}^{\sigma]}\,.
\end{align}

\begin{figure}
	\centering
	\includegraphics[width=0.57\textwidth]{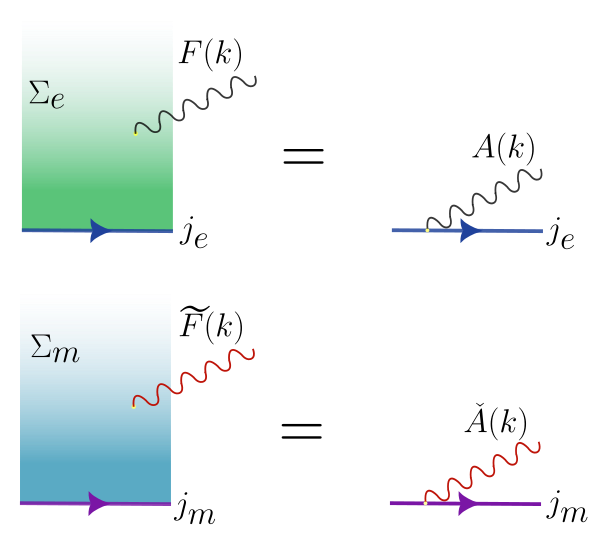}
	\caption{Photon-source vertex in field strength based perturbation theory is mathematically equal to the vertex in the new potential based formalism.}
	\label{fig:F_perturb}
\end{figure}

As shown in \cite{Aggarwal:2025fiq}, the two-form sources $\Sigma_m$ and $\Sigma_e$ appearing in the field-strength based formulation map directly onto the familiar one-form currents in the potential-based picture. The source $\Sigma_m$ , which couples to the dual field strength $\widetilde{F}$, corresponds to the magnetic current $j_m$ responsible for emitting or absorbing photons with field strength $\widetilde{F}$. Similarly, $\Sigma_e$ corresponds to the electric current $j_e$, which emits or absorbs photons carrying the field strength $F$. This follows from the action \eqref{eq:old-action} and the 2-form source and current identifications in \pref{eq:current-2-form} (see \cref{fig:F_perturb}).

Thus it is clear that in the field strength based perturbation theory, if we expand in $e$ we need to use the $\langle FF\rangle$ correlator for sourcing photon with field strength $F$ and if we expand in $g$ then we use the $\langle \widetilde{F}\widetilde{F}\rangle$ correlator for sourcing photon with field strength $\widetilde{F}$.
In the new potential based formalism, we will show that we recover the exact field strength two point functions of \cite{Aggarwal:2025fiq} below. 

To see this, we go back to the action \eqref{eq:old-action} but now written in the potential formalism \eqref{eq:sen-potential-action}. Since the 1-form currents $j_e$ and $j_m$ are defined with the electric ($e$) and magnetic ($g$) couplings respectively, we first rescale the gauge fields as
\begin{align}
    A^\mu \rightarrow \frac{1}{e} A^\mu \,, \quad \check{A}^\mu \rightarrow \frac{1}{g} \check{A}^\mu\,,
\end{align}
so that the matter interaction terms to the gauge fields 
\begin{align}
   \left( g\phi^\dagger \overleftrightarrow{\partial}_\mu \phi \check{A}^\mu + e\bar{\psi}\gamma_\mu \psi A^\mu \right)\rightarrow \left( \phi^\dagger \overleftrightarrow{\partial}_\mu \phi \check{A}^\mu + \bar{\psi}\gamma_\mu \psi A^\mu \right)
\end{align}
and matter kinetic terms with gauge covariant derivatives become coupling constants independent.
Since, in the field strength based perturbation theory, $F$ and $\widetilde{F} $ are sourced by the electric and magnetic charges, respectively (see \cref{fig:F_perturb}), we can turn on either the $\langle AA\rangle$ or the $\langle \check{A}\check{A}\rangle$ two point functions respectively. Thus, with the above scaling of the gauge fields we get
\begin{align}
   \langle F_{\mu \nu}(k)F^{\rho \sigma}(k')\rangle &=\frac{i}{8e^2}\delta^{(4)}(k+k')\frac{1}{k^2} k_{[\mu}k^{[\rho} \delta_{\nu]}{}^{\sigma]} \,,\cr
    \langle \widetilde{F}_{\mu \nu}(k)\widetilde{F}^{\rho \sigma}(k')\rangle &=\frac{i}{8g^2}\delta^{(4)}(k+k')\frac{1}{k^2} k_{[\mu}k^{[\rho} \delta_{\nu]}{}^{\sigma]}\,,
\end{align}
 recovering the two-point functions obtained in \cite{Aggarwal:2025fiq}.
 On the other hand, $\langle F\widetilde{F}\rangle$ can be found from either of the above two point functions
\begin{align}
    \langle F_{\mu \nu}(k) \widetilde{F}_{\alpha \beta}(k') \rangle=\frac{1}{2}\epsilon_{\alpha\beta \rho \sigma} \langle F_{\mu \nu}(k)F^{\rho \sigma}(k')\rangle =-\frac{i}{8 e^2 k^2}\delta^{(4)}(k+k')k^{[\mu}\epsilon^{\nu] \alpha \beta \gamma}k_\gamma \,.
\end{align}
This match the two point function of \cite{Aggarwal:2025fiq} up to an overall normalisation factor. This completes the demonstration that both Sen's formalism and our new action lead to identical perturbative results via Feynman diagrams.

\subsection{Supersymmetric QEMD in \texorpdfstring{$D=4$}{D=4}}

Including matter, we shall consider the following non-supersymmetric action
\begin{align}\label{action_non_susy}
    S=S[{\mathcal{B}},\check{\mathcal{B}}]+\int d^4x \Big[-\frac{1}{2} \partial_\mu A_{\nu}\partial^\mu A^\nu -\frac{1}{2} \partial_\mu \check{A}_{\nu}\partial^\mu \check{A}^\nu  - |\check{D}_\mu \phi|^2-i\bar{\psi}\slashed{D}\psi
   -m_\phi^2 |\phi|^2-m_\psi \bar{\psi}\psi \Big]\,,
\end{align}
where
\begin{align}
    S[{\mathcal{B}},\check{\mathcal{B}}]=\frac{1}{2}\int \d \mathcal{B} \wedge \star \d \mathcal{B} +\frac{1}{2}\int \d \check{\mathcal{B}} \wedge \star \d \check{\mathcal{B}}\,,
\end{align}
with the additional fields defined as: $\mathcal{B}:=B-A \,; \check{\mathcal{B}}:=\check{B}+\check{A}$ 
and for the matter sector, we consider Dirac spinor and complex scalar field coupled with the gauge fields $A^\mu$ and $\check{A}^\mu$ respectively. Here the gauge covariant derivatives are defined as
\begin{align}
    \check{D}_\mu=\partial_\mu +i \check{A}_\mu \,, \qquad \slashed{D}=\slashed{\partial}+i \slashed{A}\,,
\end{align}
where we have absorbed the gauge couplings into the gauge fields: ($g\check{A}\rightarrow \check{A},eA \rightarrow A$). The full supersymmetric action is given by
\begin{align}
    S=S_{\text{gauge}}+ S_{\text{matter}}\,,
\end{align}
where
\begin{align}\label{susy_actions}
    S_{\text{gauge}} &=-\frac{1}{4}\int d^4 x \left[ \int d^2\theta  W_{i,\alpha} W_{i}^\alpha +\int d^2\bar{\theta}  \overline{W}_{i,\dot{\alpha}} \overline{W}_{i}^{\dot{\alpha}} \right]\,,\cr
    S_{\text{matter}} &=\int d^4 x \int d^2\theta d^2\bar{\theta} \left[\bar{\Phi}_1e^{V_1} \Phi_1 +\Phi_2 e^{-V_1}\bar{\Phi}_2 + \bar{\tilde{\Phi}}e^{V_2}\tilde{\Phi}\right]\,,
\end{align}
 and the supersymmetry transformations for the component fields are given by 
 \begin{align}\label{susy_transofrmation_comp}
     \delta_S A_{\mu} &=i \bar{S}\bar{\sigma}_\mu \lambda_1-i\bar{\lambda}_1\bar{\sigma}_\mu S\,,\cr
      \delta_S \check{A}_{\mu} &=i \bar{S}\bar{\sigma}_\mu \lambda_2-i\bar{\lambda}_2\bar{\sigma}_\mu S\,.
 \end{align}
The definitions of various superfields and there transformations can be found in any standard reference on supersymmetry (see, for example, \cite{ArgurioSUSYNotes, Wess:1992cp}). Below, we only spell out how the fields in non supersymmetric action are packaged into these superfields.
 The chiral superfields appearing in S$_{\text{gauge}}$ are defined
 \begin{align}\label{chiral_superfield}
    W_{i,\alpha}:=-\frac{i}{4}\overline{\D}^2 \D_\alpha V_i\,, \qquad \overline{W}_{i,\dot{\alpha}}=-\frac{i}{4}\D^2 \overline{\D}_{\dot{\alpha}} V_i\,,\qquad i=1,2\,,
\end{align}
where the real superfields $V_i(x,\theta,\bar{\theta})$ in the Wess-Zumino gauge are given in terms of the following component fields
\begin{align}
    V_1(x,\theta,\bar{\theta}) &=\theta \sigma^\mu \bar{\theta}A_{\mu} +i\theta^2\bar{\theta}\bar{\lambda}_1-i\bar{\theta}^2\theta \lambda_1+\frac{1}{2}\theta^2 \bar{\theta}^2 d_1\,,\cr
    V_2(x,\theta,\bar{\theta}) &=\theta \sigma^\mu \bar{\theta}\check{A}_{\mu} +i\theta^2\bar{\theta}\bar{\lambda}_2-i\bar{\theta}^2\theta \lambda_2+\frac{1}{2}\theta^2 \bar{\theta}^2 d_2\,.
\end{align}
The lowest components are the U(1) gauge fields $A^\mu$ and $\check{A}^\mu$ -present in the non-supersymmetric action, ($\lambda_{i,\alpha},\bar{\lambda}_{i,\Dot{\alpha}}$) are Weyl fermions also known as gaugino and $d_i$ are real auxiliary scalar fields.
The chiral superfields in S$_{\text{matter}}$ are defined in terms of component fields as
\begin{align}\label{matter_superfields}
    \Phi_i &=\rho_i +\sqrt{2}\theta \chi_i +i\theta \sigma^\mu \bar{\theta}\partial_\mu \rho_i +\theta^2 f_i +\frac{i}{\sqrt{2}}\theta^2 \bar{\theta}\bar{\sigma}^\mu \partial_\mu \chi_i -\frac{1}{4}\theta^2 \bar{\theta}^2 \Box \rho_i\,,\cr
    \tilde{\Phi}&=\phi +\sqrt{2}\theta \xi +i\theta \sigma^\mu \bar{\theta}\partial_\mu \phi +\theta^2 \tilde{f}_i +\frac{i}{\sqrt{2}}\theta^2 \bar{\theta}\bar{\sigma}^\mu \partial_\mu \xi -\frac{1}{4}\theta^2 \bar{\theta}^2 \Box \phi\,.
\end{align}
where $i=1,2$ and we package both the Weyl spinors $\chi_i$ into a Dirac spinor $\psi$
\begin{align}\label{Dirac_fermion}
    \psi =\begin{pmatrix}
        \chi_{1\alpha}\\
        \bar{\chi}_{2}^{\dot{\alpha}}
    \end{pmatrix}\,,
\end{align}
where $\bar{\chi}_{2}^{\dot{\alpha}}$ is in the conjugate representation of $\chi_{1\alpha}$.

 For the extra fields $B^\mu$ and $\check{B}^\mu$, as discussed earlier (see \eqref{susy_B}), we simply set the supersymmetry transformations for $B_i^\mu$ fields as
\begin{align}
    \delta_S B^\mu &=\delta_S A^\mu=i\bar{S}\bar{\sigma}_\mu \lambda_1-i\bar{\lambda}_1\bar{\sigma}_\mu S\,, \cr
    \delta_S \check{B}^\mu &=-\delta_S \check{A}^\mu=-i\bar{S}\bar{\sigma}_\mu \lambda_2+i\bar{\lambda}_2\bar{\sigma}_\mu S \,,
\end{align}
using \eqref{susy_transofrmation_comp}. 
Although we have supersymmetrised the action in \eqref{action_non_susy}, it can be easily checked that the supersymmetric version of the original action in \eqref{action_total} is the same action in \eqref{susy_actions} since the field redefinitions from $(\mathcal{B},\check{\mathcal{B}})$ to $(B,\check{B})$ respect the above supersymmetry transformations.

\section{Discussion} \label{sec:disc}
The central result of this paper is a new action that can handle duality invariant QFTs in arbitrary dimensions. The new action crucially relies on extending Sen's action that introduces a potential and a specific higher-form gauge symmetry, which we dubbed $h$-gauge symmetry. This symmetry can be gauge fixed to lead us back to Sen's action, or equivalently to one involving only the potential. The potential based action is written in terms of a shadow sector, which can be trivially integrated out to result into familiar Maxwell-like actions for the gauge fields. However, since this action is now a gauge-fixed version of a parent action, the definition of the field-strength is inherited from the parent theory. There is a higher-form residual symmetry that keeps track of this \textit{correct} field-strength which we called harmonic symmetry.  We performed a number of non-trivial consistency check and described how our action subsumes and supersedes Sen's string theoretic formalism.

Given such a decoupling, it is prudent to ask what role, if any, are played by the shadow sector in our formalism once we choose a gauge fixing that keeps only the potentials. In flat spacetime, they can indeed be happily ignored as long as one keeps track of the harmonic symmetry that fixes the correct definition of the field-strength. However, to couple to dynamic gravity, the correct starting point in flat spacetime is \textit{not} the usual Maxwell-like actions. Let us explain why. For specificity, let us focus on the case of self-dual fields, the extension to the general case is immediate. 

Suppose we start with self-dual fields in flat spacetime described by the Maxwell-like action (but with the correct definition of the field-strength) and minimally couple it to gravity.

\begin{equation}
    S = - \frac{1}{2} \int_{\mathbb{R}^{(1,4n+1)}} \d C \wedge \star \d C \xrightarrow{\star \,\to \, \star_g} - \frac{1}{2} \int_{\mathcal{M}} \d C \wedge \star_g \d C \;.
\end{equation}
Under such a minimal coupling, the stress-tensor will now receive contribution from the minimally coupled self-dual field-strength $F_g = \d C + \star_g \d C$ but also from the anti-self-dual field strength $F^-_g = \d C - \star_g \d C$. Therefore, after minimal coupling, the theory is no longer describing the dynamics of a self-dual field alone. 

In general, this is expected and known that for duality invariant theories, the gravitational coupling needs to be non-standard, see \citep{Witten:2009at, Lambert:2019diy}. The exact same predicament exists for Sen's formulation as well. The solution is conceptually straightforward --- treat coupling to gravity as any other interactions and continue to write the kinetic terms using the shadow sector. This map for self-dual case is explicitly worked out in \citep{Sen:2015nph,Sen:2019qit, Andriolo:2020ykk,Hull:2023dgp} and has been subjected to a number of non-trivial tests \citep{Chakrabarti:2020dhv,Chakrabarti:2022jcb,Chakrabarti:2023czz,Vanichchapongjaroen:2025psm,Hull:2025rxy}. The solution is exactly same for our new formalism, with one important extension, unlike Sen's, our construction needs to work for any general dimensions. The details of the construction will be reported elsewhere \citep{future}.

 Also note, that our construction can also easily accommodate non-linear self-dual electrodynamics that recently has seen a revival in the community (see \citep{Sorokin:2021tge} and references therein). Essentially, if one considers the following scalars $s,\check{s}, p, \check{p}$ defined as (in $D=4$)

\begin{align}
    \star s &= \d A \wedge \star \d A \;, \; \star \check{s} = \d \check{A} \wedge \star \d \check{A} \nonumber \\
    \star p & = \d A \wedge \d A \;, \; \star \check{p} = \d \check{A} \wedge  \d \check{A} \;,
\end{align}
then any Lagrangian function $\mathcal{L}(s,p,\check{s},\check{p})$ that is invariant under $(s,p) \leftrightarrow (\check{s},\check{p})$ will be duality invariant.

\section*{Acknowledgement}
SC thanks Ashoke Sen for discussions and substantial feedback on an earlier version of this draft.
SC thanks IMSc and the organisers and participants of the conference ``Quantum fields and strings" held at IMSc, Chennai where some of the results reported here were presented. MR thanks Ashoka University and the participants of the enjoyable two-day symposium on theoretical physics, where an earnest effort to present \emph{all} of these results was made. SC acknowledge the financial support by the Czech Science Foundation (GA\v{C}R) grant “Dualities and higher derivatives” (GA23-06498S). MR is supported by an Inspire Faculty Fellowship.

\bibliography{refs}
\end{document}